\documentclass[12pt]{article}
\usepackage{amsmath,amsfonts,amssymb,amsthm,bbm,algorithm}
\usepackage{algpseudocode}
\usepackage{booktabs}
\usepackage{multirow,threeparttable}
\usepackage{natbib,enumerate}
\usepackage{setspace}
\setcitestyle{authoryear,open={(},close={)}} 
\usepackage{authblk} 
\usepackage[prependcaption,colorinlistoftodos,textsize=small]{todonotes}

\usepackage{geometry}
\usepackage{amsmath,amsthm,amssymb,marvosym,enumitem,makeidx,hyperref,graphicx,xcolor,enumerate,soul,verbatim}
\renewcommand{\le}{\leqslant}  

\newcommand{\norm}[1]{\left\lVert #1 \right\rVert}

\def\*#1{\mathbf{#1}} 
\def\set#1{\left\{ {#1} \right\}} 

\def\bR{{\mathbb R}}

\def\sC{{\mathcal C}}

\def\sE{{\mathcal E}}

\def\sI{{\mathcal I}}

\def\sL{{\mathcal L}}
\def\sM{{\mathcal M}}
\def\sN{{\mathcal N}}

\def\sS{{\mathcal S}}
\def\sT{{\mathcal T}}

\theoremstyle{plain}
\newtheorem{theorem}{Theorem}[section]

\newtheorem{lemma}[theorem]{Lemma}

\theoremstyle{remark}

\def\tr{{\text{tr}}}

\title{Bayesian covariance regression for differential network analysis of zero-inflated microbiome data}

\author[1]{Zichun Xu}
\author[2]{Jing Ma\thanks{Corresponding author: jingma@fredhutch.org}}

\affil[1]{Department of Biostatistics, University of Washington}
\affil[2]{Division  of Public Health Sciences, Fred Hutchinson Cancer Center}

\begin{document}
\maketitle

\begin{abstract}
Microbial interaction networks can rewire in response to host and environmental factors, yet most existing methods for network estimation treat the covariance structure as static across samples. We propose TRECOR, a Bayesian covariance regression framework for inferring covariate-dependent microbial covariation networks from zero-inflated compositional count data. The method models microbiome counts through a latent multivariate normal distribution defined on the internal nodes of a phylogenetic tree, where both the mean and covariance of the latent variables depend on covariates. The covariance is decomposed into a sparse baseline component, representing a stable microbial covariation network, and a low-rank covariate-dependent perturbation that captures network rewiring. By exploiting the binomial factorization of the multinomial distribution under the logistic-tree-normal representation, the model achieves full conjugacy and posterior inference proceeds via an efficient Gibbs sampler. In simulations, TRECOR substantially outperforms covariance regression applied to transformed counts, demonstrating the importance of explicitly modeling the compositional sampling layer. Applied to gut microbiome data from 531 individuals across three countries, we find that age has the largest effect on microbial covariation, a pattern not revealed by mean-based analysis alone. The age-associated differential network is enriched for \textit{Enterobacteriaceae} and related families, consistent with known developmental shifts in the gut microbiota, while country-associated differential networks implicate diet-related taxa.
\end{abstract}

\newpage
\doublespacing
\section{Introduction}

The study of microbial covariation networks is essential for understanding the structure and function of ecological systems \citep{faust2012microbial}. While many methods exist to infer these networks, most assume that the underlying network structure is static across samples \citep{friedman2012inferring,fang2015cclasso,yoon2019microbial,lin2022linear,deek2023inference}. In reality, microbial interaction networks can undergo rewiring in response to environmental factors or host conditions, reflecting a fundamental and dynamic property of microbiome ecosystems \citep{chen2020gut}. Identifying these covariate-dependent changes is critical for the development of microbiome-based precision interventions. In this work, we aim to infer covariate-dependent microbial covariation networks, where the covariate may be binary, continuous, or categorical. 

\subsection{Motivating study}
Our work is motivated by a human microbiome study consisting of 531 individuals across ages, sexes, and geographic regions \citep{yatsunenko2012human}. While previous analyses of this cohort established significant variation in microbial abundances across these factors \citep{zeng2021model,wang2023generalized}, how these factors influence microbial covariation remains largely unexplored.

\begin{figure}[h!]
    \centering
    \includegraphics[width=0.9\linewidth]{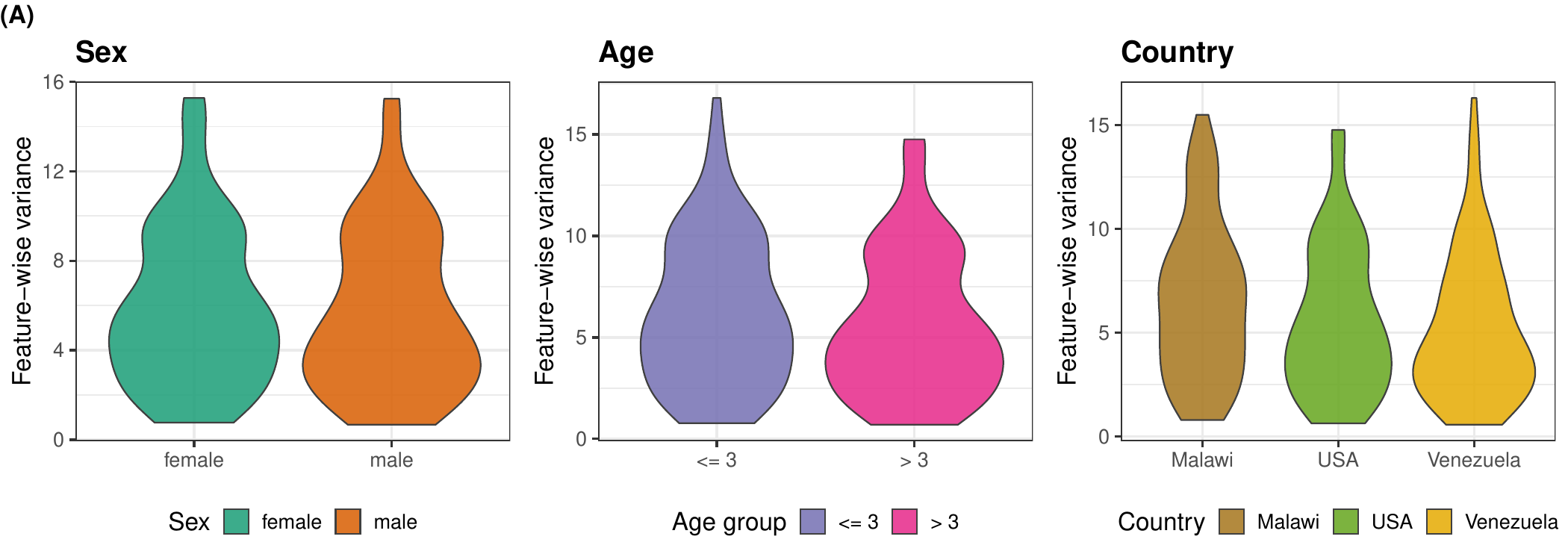}
    \includegraphics[width=0.9\linewidth]{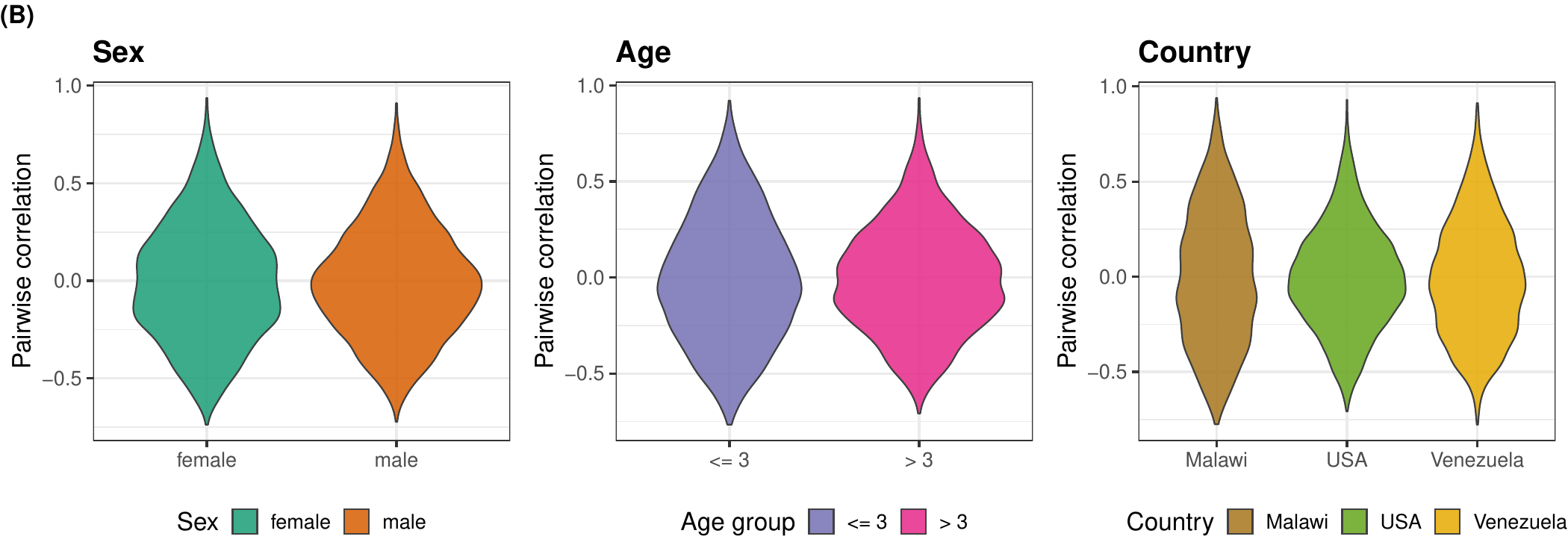}
    \caption{Illustration of covariate-dependent covariation in the \cite{yatsunenko2012human} study. (A) Feature variances; (B) Feature correlations.}
    \label{fig:covariation}
\end{figure}

As an illustration, we applied the centered log-ratio transformation to pseudocount-adjusted counts and examined the distributions of taxon-wise variances and pairwise correlations across groups defined by sex, age, and country of origin. As shown in Figure \ref{fig:covariation}, microbial covariation patterns differ markedly across countries and, to a lesser extent, across age groups. These observations suggest that host and environmental factors influence not only the abundance of individual taxa but also the covariation structure among taxa. Capturing this ``rewiring" requires a statistical framework that allows the covariance structure, rather than only the mean vector, to depend on covariates.

While Figure~\ref{fig:covariation} displays covariation at the genus level, an equivalent signal is observable at the level of internal nodes of the phylogenetic tree, which aggregate related genera in a hierarchical manner (Supplementary Figure~\ref{fig:covariation:internalnodes}). As discussed in Section~\ref{sec:methods}, operating at the internal-node level offers important statistical advantages, including mitigating zero inflation and improved computational tractability, while preserving the same biological signal.

Covariance regression for Gaussian data has been extensively studied \citep{hoff2012covariance,fox2015bayesian,kim2025high,auddy2024regressing}. However, these frameworks are not well suited for microbiome data because they do not account for two key characteristics: zero inflation and the compositional constraint imposed by the uneven library sizes. Recent work such as MiCoRe \citep{mcgregor2020bayesian} attempts to address these challenges by modeling compositional counts using the logistic-normal-multinomial (LNM) distribution within a Bayesian covariance regression framework. However, the lack of conjugacy between the multinomial sampling model and the logistic-normal distribution makes the LNM computationally challenging in high dimensions. In practice, \cite{mcgregor2020bayesian} applied MiCoRe only to the 25 most abundant families in their dataset and noted that the method does not scale well to larger numbers of taxa. Moreover, the rank-one perturbation assumed by MiCoRe imposes a strong constraint on how covariates influence the covariance matrix, which may be insufficient to capture heterogeneous covariate effects commonly observed in microbiome data. 

\subsection{Our contributions}
In this paper, we propose a Bayesian covariance regression framework for zero-inflated compositional count data based on the logistic-tree-normal (LTN) distribution \citep{wang2021tree,leblanc2023microbiome}. Our model represents microbiome counts through a latent multivariate normal distribution defined on the internal nodes of a phylogenetic tree, where both the mean and covariance of the latent variables are allowed to depend on covariates. We introduce a sparse plus low-rank decomposition of the covariance matrix that separates a stable baseline microbial covariation network from low-rank perturbations associated with covariates. Finally, by exploiting the binomial factorization of the multinomial distribution under the LTN representation, we obtain conjugate Bayesian inference via efficient Gibbs sampling, enabling scalable inference in large-scale microbiome studies.

We demonstrate that the proposed method enhances estimation of the microbial covariation networks through extensive simulation studies. When applied to the gut microbiome data from \cite{yatsunenko2012human}, we find that age has the largest effect on microbial covariation, a pattern not revealed by mean-based analysis alone. The differential correlation networks associated with age are enriched for taxa in \textit{Enterobacteriaceae} and related families, consistent with known shifts in the gut microbiota during early development. Differential networks associated with country of origin implicate diet-related taxa, reflecting dietary differences between the US and non-US cohorts. In addition, the estimated population correlation network exhibits a sparse structure concentrated among phylogenetically proximate internal nodes, suggesting that baseline microbial co-variation is primarily driven by closely related taxa.

\subsection{Related work}

Covariate-dependent networks have been studied in the context of graphical models or conditional dependence networks \citep{cheng2014sparse,ni2022bayesian,wang2025high}.
These ideas have recently been extended to the study of differential microbial conditional dependence networks associated with a binary covariate \citep{cai2019differential,he2019direct,mcgregor2020mdine}.
However, estimation and inference for conditional dependence networks are generally harder and require stronger assumptions, especially in high-dimensional settings. Conditional dependence networks are also sensitive to latent confounding. When unobserved variables (latent confounders) influence multiple observed nodes in a network, they create spurious correlations that can be easily mistaken for true conditional dependencies \citep{chandrasekaran2012latent}. Given these challenges, we focus instead on inference for covariation networks, which are scientifically useful in their own right. Moreover, covariation-based analyses are more stable and require fewer modeling assumptions when sample sizes are limited and many taxa are rare.

\section{Methods}\label{sec:methods}

\textit{Notations.} Throughout, bold letters denote vectors and matrices. For any positive integer $m$, let $[m]=\{1,\dots,m\}$. We write $(\cdot)^\top$ for transpose, $\text{diag}(\cdot)$ for the diagonal matrix formed from the diagonal entries of its argument, and $\|\cdot\|_2$ and $\|\cdot\|_F$ for the Euclidean and Frobenius norms, respectively. Let $\mathbb{R}^d$ denote the $d$-dimensional Euclidean space and $\mathcal{N}_q(\boldsymbol{\mu},\*\Sigma)$ denote the $q$-variate normal distribution with mean $\boldsymbol{\mu}$ and covariance matrix $\*\Sigma$. 

\subsection{The logistic-tree model}
Consider a microbiome dataset over $n$ samples $\set{\boldsymbol{z}_i}_{i=1}^n$, where $\boldsymbol{z}_i = [z_{ij}, \ldots, z_{i(q+1)}]^\top \in \bR^{q+1}$ represents the counts of $q+1$ taxa measured in the $i$-th sample. For each sample, let $M_i = \sum_{j=1}^{q+1} z_{ij}$ denote its library size and $\boldsymbol{x}_i \in \bR^d$ denote a vector of $d$ covariates. We assume that the number of taxa $q+1$ is large and grows with the sample size $n$, whereas the number of covariates $d$ is small or moderate. This high-dimensional setting is common in many real data applications. 

To capture the compositional constraint, the multinomial distribution is widely used to model taxa counts 
\begin{equation*}
\begin{aligned}
& \boldsymbol{z}_i\mid \boldsymbol{\theta}_i \sim \text{Multinomial}(M_{i}, \boldsymbol{\theta}_i),
\end{aligned}
\end{equation*}
where the latent relative abundances $\boldsymbol{\theta}_i = [\theta_{i1}, \ldots, \theta_{i(q+1)}] \in \bR^{q+1}$ are further modeled by some probability distribution, e.g., the Dirichlet or logistic-normal distribution. 
However, the multinomial sampling model coupled with common choices for \(\boldsymbol{\theta}_i\) can be either too restrictive to capture complex taxon-taxon dependence or computationally challenging to fit in high dimensions. 

The logistic-tree model addresses this issue by decomposing the multinomial sampling model into a series of independent binomial sampling models \citep{wang2021tree,leblanc2023microbiome}. More specifically, let $\sT = \sT(\sI, \sL; \sE)$ represent a binary phylogenetic tree over $q+1$ taxa, where $\sI$, $\sL$, and $\sE$ are the set of internal nodes, leaf nodes, and edges of the tree $\sT$, respectively. Then, $|\sI| = q$ and $|\sL| = q+1$. Let $j \in \sI$ denote any internal node of $\sT$, and let $\sC(j) \subset \sL$ and $\sC_L(j) \subset \sL$ denote the set of leaf nodes corresponding to the sub-tree and left sub-tree of $j$, respectively. Define 
\begin{equation}
\label{eq:internal_counts}
p_{ij} = \frac{\sum_{k \in \sC_L(j)} \theta_{ik}}{\sum_{k \in \sC(j)} \theta_{ik}}, \quad N_{ij} = \sum_{k \in \sC(j)}z_{ik}, \quad y_{ij} = \sum_{k \in \sC_L(j)}z_{ik},    
\end{equation}
and $\boldsymbol{p}_i = [p_{i1}, \ldots, p_{iq}]^\top$. It is not hard to verify that the map between $\boldsymbol{\theta}_i$ and $\boldsymbol{p}_i$ is bijective. Let \(\varphi_{ij}\) denote the logit transformation of $p_{ij}$ and define $\boldsymbol{\varphi}_i = [\varphi_{i1}, \ldots, \varphi_{iq}]^\top$. The logistic-tree model assumes that the data follow the distribution
\begin{equation}
\begin{aligned}
\label{eq:expression_measurement_2}
 \boldsymbol{\varphi}_i & \sim G_{q}(\boldsymbol{x}_i), \quad i = 1, \ldots, n\\
 \varphi_{ij} & = \log\left(\frac{p_{ij}}{1-p_{ij}}\right), \quad j = 1, \ldots, q\\
 y_{ij}\mid p_{ij} & \sim \text{Binomial}(N_{ij}, p_{ij}),
\end{aligned}
\end{equation}
where $G_{q}(\boldsymbol{x})$ is some probability distribution over $\bR^{q}$ that depends on the covariates in $\boldsymbol{x}$.

Each latent variable $\varphi_{ij}$ encodes a \emph{clade-level compositional balance}: specifically, the log-odds that a read drawn uniformly from clade $\mathcal{C}(j)$ in sample $i$ belongs to the left sub-clade $\mathcal{C}_L(j)$ rather than the right. Intuitively, $\varphi_{ij}$ describes how the total abundance within clade $j$ is partitioned between two groups of related taxa. A sample with a large positive $\varphi_{ij}$ has the left sub-clade dominant within that clade, regardless of the clade's total abundance.


\subsection{Biological interpretation of internal-node covariation}

Covariation between two internal nodes $j$ and $k$ measures whether these within-clade partitionings tend to shift together across samples. Positive covariation indicates that samples skewed toward the left sub-clade at $j$ are also skewed at $k$, reflecting coordinated restructuring across two groups of related taxa. Differential covariation, i.e., a change in the covariance of $(\varphi_{ij}, \varphi_{ik})$ with respect to a covariate, therefore indicates that this coordination is reshaped by that factor. A high-degree internal node in a differential covariation network identifies a clade whose within-group balance is most broadly coupled to and most perturbed by community-wide compositional shifts. The leaf taxa descending from it provide an interpretable biological summary of the taxa most affected by the covariate.

This internal-node representation differs from leaf-level taxon covariation, and the transformation between them is nonlinear \citep{greenacre2020amalgamations}. However, internal-node covariation preserves the same biological signal (Supplementary Figure~S4) while offering two key statistical advantages: clade-level counts are far less zero-inflated than leaf-level counts, and the binomial factorization yields a fully conjugate model that enables efficient Gibbs sampling. As discussed in \cite{wang2021tree}, counts drawn from the logistic-tree sampling model coupled with a multivariate normal distribution on \( \boldsymbol{\varphi}_i\) exhibit features similar to those observed in empirical data, including excess of zeros. 

\subsection{A covariance regression model}
We now detail the model $G_{q}(\boldsymbol{x})$. In the literature, association between microbial abundance and covariates has been mostly restricted to the mean abundance \citep{clark2017generalized,mcgregor2020mdine,osborne2022latent}. However, co-variation among taxa, which is characterized by the covariance matrix, can also be attributed to various environmental and clinical factors \citep{chen2020gut,deek2023inference}. Motivated by the need to perform differential network analysis, we propose a flexible model that captures the effect of covariates on both the mean and the covariance. Specifically, we assume that each $\boldsymbol{\varphi}_i$ is sampled from the following multivariate normal distribution:
\begin{equation}
\label{eq:cov_reg_model}
\boldsymbol{\varphi}_i \sim \mathcal{N}_{q}\!\left(
\mathbf{B}_0 \boldsymbol{x}_i,\;
{\*\Sigma} + \sum_{r=1}^R \mathbf{B}_r \boldsymbol{x}_i \boldsymbol{x}_i^{\top} \mathbf{B}_r^{\top}
\right).
\end{equation}
Here, both the mean and the covariance matrix depend on  covariates, with unknown parameters $\*\Sigma$ and $\set{\mathbf{B}_r \in \bR^{q \times d}}_{r=0}^R$. The baseline covariance matrix $\*\Sigma$ is positive definite and assumed to be sparse. The covariate-dependent part of the covariance $\sum_{r=1}^R\*B_r \boldsymbol{x}_i \boldsymbol{x}_i^\top\*B_r^\top$ has rank $R$ with $R \ll q$. Therefore, the expression model \eqref{eq:cov_reg_model} is a generalization of the ``sparse plus low-rank" structure \citep{bertsimas2023sparse} to the case where the low rank part is covariate-dependent. The set of parameters $\set{\*B_r}_{r=1}^R$ thus characterizes the effect of covariates on the covariance matrix. However, these parameters are not identifiable. For example, flipping the sign of any $\*B_r$ clearly leads to an equivalent model. Lemma~\ref{lem:identifiablity} establishes the identifiability of model \eqref{eq:cov_reg_model}. 
\begin{lemma}
\label{lem:identifiablity}
Define: $$
\mathbf{B}^{(j)} \in \bR^{q \times R} \equiv \big(\boldsymbol{b}_{j1}, \ldots , \boldsymbol{b}_{jR}\big),\quad j \in [d]$$
where $\boldsymbol{b}_{jr} \in \bR^{q}$ is the $j$-th column of $\*B_r$. The set of parameters $\set{\*B_r}_{r=1}^R$ can thus be equivalently represented as $\set{\mathbf{B}^{(j)}}_{j=1}^d$. Suppose $\set{\*A_r}_{r=1}^R$ is another set of parameters with equivalent representation $\set{\*A^{(j)}}_{j=1}^{d}$. Then $\sum_{r=1}^R\*B_r\boldsymbol{x} \boldsymbol{x} ^\top\*B_r^\top \equiv \sum_{r=1}^R\*A_r\boldsymbol{x}\boldsymbol{x}^\top\*A_r^\top$ for any $\boldsymbol{x} \in \bR^{d}$ if and only if there exists an orthonormal matrix $\*Q \in \bR^{R \times R}$ such that $\*A^{(j)} = \mathbf{B}^{(j)}\*Q$ for all $j \in [d]$.
\end{lemma}
Lemma~\ref{lem:identifiablity} says that model \eqref{eq:cov_reg_model} is identifiable up to a rotation, represented by an $R \times R$ orthonormal matrix $\*Q \in \bR^{R \times R}$. This has a direct consequence for inference: individual matrices $\*B_r$ are not interpretable on their own, as they can be arbitrarily rotated without changing the likelihood. Only rotation-invariant quantities are meaningful. In particular, for each covariate $j \in [d]$, the Frobenius norm $\norm{\*B^{(j)}}_F$ is invariant to the rotation $\*Q$ and thus identifiable. We therefore use $\norm{\*B^{(j)}}_F$ as a measure of the effect of the $j$-th covariate on the covariance structure. We also note that model \eqref{eq:cov_reg_model} can be equivalently written as the following factor model:
\begin{equation*}
\begin{aligned}
\boldsymbol{\gamma}_i & \sim \mathcal{N}_{R}(\boldsymbol{0}, \*I_R)\\
\boldsymbol{\varphi}_i\mid\boldsymbol{\gamma}_i & \sim \mathcal{N}_{q}\left(\mathbf{B}_0 \boldsymbol{x}_i+\sum_{r=1}^R\gamma_{ir}\*B_r\boldsymbol{x}_i, \*\Sigma\right),
\end{aligned}
\end{equation*}
where $\boldsymbol{\gamma}_i = [\gamma_{i1},\ldots,\gamma_{iR}]^{\top}$ are latent factors with covariate-dependent loading matrices $\set{\*B_r \boldsymbol{x}_i}_{r=1}^R$. 

The model in \eqref{eq:cov_reg_model} builds upon the linear covariance regression framework first introduced by \cite{hoff2012covariance}. While \cite{mcgregor2020bayesian} recently extended this approach to sparse count data within a Bayesian framework, both of these existing methods are restricted to rank-one ($R = 1$) perturbations. Such a constraint is often insufficient for capturing the complex, heterogeneous covariate effects observed in biological systems. Furthermore, these earlier models do not scale well to high-dimensional settings, where the number of taxa is large. Our approach addresses these gaps by allowing for an arbitrary rank $R$ and utilizing the full conjugacy of the LTN distribution to handle high-dimensional microbiome data. 

\subsection{Priors}
To complete the model, we now specify the priors for parameters $\*\Sigma$ and $\set{\*B_r \in \bR^{q \times d}}_{r=0}^R$. Due to the high dimensionality, we consider the graphical LASSO prior for $\*\Sigma$ \citep{wang2012bayesian}:
\begin{equation*}
\begin{aligned}
&\pi(\*\Sigma\mid \boldsymbol{\tau}, \lambda) \propto C_{\boldsymbol{\tau}}^{-1}\prod_{i<j}\set{\text{Lap}(\sigma_{ij}\mid\boldsymbol{\tau}_{ij})}\prod_{j=1}^q\set{\text{Exp}\left(\sigma_{jj}\mid\frac{\lambda}{2}\right)}\*1\set{\*\Sigma \succ 0}, \\
&\pi(\boldsymbol{\tau}\mid \lambda) \propto C_{\boldsymbol{\tau}}\prod_{i < j }\text{Exp}\left(\tau_{ij}\mid \frac{\lambda^2}{2}\right),
\end{aligned}
\end{equation*}
where $C_{\boldsymbol{\tau}} > 0$ is some normalizing constant that depends on $\boldsymbol{\tau}$, 
$\sigma_{ij}$ is the $(i,j)$-th entry of $\*\Sigma$, and $\text{Lap}(\cdot\mid\lambda)$ and $\text{Exp}(\cdot\mid\lambda)$ represent a Laplace distribution and an exponential distribution with parameter $\lambda$, respectively. We assign a gamma prior to the shrinkage parameter $\lambda$: 
\begin{equation*}
\pi(\lambda\mid v,s) = \text{Gamma}(v,s).
\end{equation*}
To specify a prior for $\set{\*B_r}_{r=1}^R$, it is important to take into account the identifiability condition established in Lemma~\ref{lem:identifiablity}. For any two sets of parameters that lead to an equivalent model, an appropriate prior should place equal weights on them. Motivated by this, we consider placing independent normal priors on each column of $\*B_r$
\begin{equation*}
\pi(\boldsymbol{b}_{jr}\mid \*\Sigma,\nu_j) = \mathcal{N}(\boldsymbol{0},\*\Sigma\nu_j),
\end{equation*}
for all $j \in [d]$ and $r \in [R]$. The priors of $\set{\*B_r}_{r=1}^R$ are thus
\begin{equation}
\label{eq:priors_B}
\begin{aligned}
\pi\left(\set{\*B_r}_{r=1}^R\mid \*\Sigma, \*D_\nu\right) 
&\propto  \prod_{r=1}^R\exp\set{-\frac{1}{2}\text{tr}\Big(\*B_r^\top\*\Sigma^{-1}\*B_r\*D_\nu^{-1}\Big)},
\end{aligned}
\end{equation}
where $\*D_\nu \in \bR^{d \times d}$ is a diagonal matrix whose $j$-th diagonal element is $\nu_j$ and \(\mbox{tr}(\cdot)\) is the trace operator. Equivalently, we have independent matrix normal priors for \(\mathbf{B}_r\)
\begin{equation}
\notag
\pi(\*B_r\mid \*\Sigma, \*D_\nu) = \mathcal{MN}_{q \times d}(\boldsymbol{0}, \*\Sigma, \*D_\nu),
\end{equation}
where $\mathcal{MN}_{q \times d}(\*A, \*B, \*C)$ represents a matrix normal distribution with mean $\*A \in \bR^{q \times d}$, row covariance $\*B \in \sS_+^{q}$, and column covariance $\*C \in \sS_+^d$. 
The prior distribution  \eqref{eq:priors_B} is appropriate with respect to the identifiability condition of Lemma~\ref{lem:identifiablity}. The scale parameters $\set{\nu_j}_{j=1}^d$ reflect the effect size of the $j$-th covariate on the covariance matrix. To adaptively learn them from the data, we assign independent inverse-gamma priors on $\set{\nu_j}_{j=1}^d$,
\begin{equation*}
\pi(\nu_j\mid a_\nu, b_\nu) = \text{IG}(a_\nu, b_\nu).
\end{equation*}
For the mean effect $\mathbf{B}_0$, we consider a similar prior structure to $\*B_r$,
\begin{equation*}
\pi(\mathbf{B}_0\mid \*\Sigma, \*D_\omega) = \mathcal{MN}_{q \times d}(\boldsymbol{0}, \*\Sigma, \*D_\omega),
\end{equation*}
where $\*D_\omega \in \bR^{d \times d}$ is a diagonal matrix whose $j$-th diagonal element is $\omega_j$ and the priors on $\set{\omega_j}_{j=1}^d$ are:
\begin{equation*}
\pi(\omega_j) = \text{IG}(a_\omega, b_\omega).
\end{equation*}

\subsection{Posterior inference via Gibbs samplers}
\label{subsec:gibbs}
With the above choices of priors, the model \eqref{eq:expression_measurement_2}-\eqref{eq:cov_reg_model} achieves full conjugacy and the posterior can be sampled via a Gibbs algorithm. We summarize the key full conditional distributions below; derivations and the complete algorithm are deferred to Appendix~\ref{sec:apdx_gibbs}.


The full conditional of $\*\Sigma$ depends on the sufficient statistic
\begin{equation*}
\*S = \sum_{i=1}^n\widetilde{\boldsymbol{\varphi}}_i\left(\widetilde{\boldsymbol{\varphi}}_i\right)^\top + \sum_{r=1}^R\*B_r\*D_\nu^{-1}\*B_r^\top + \*B_0\*D_\omega^{-1}\*B_0^\top,
\end{equation*}
where $\widetilde{\boldsymbol{\varphi}}_i = \boldsymbol{\varphi}_i - \*B_0 \boldsymbol{x}_i - \sum_{r=1}^R\gamma_{ir}\*B_r \boldsymbol{x}_i$. Given $\*S$, $\*\Sigma$ is updated via an augmented block Gibbs algorithm following \cite{wang2012bayesian}.

The full conditional of $\*B_0$ is
\begin{equation}
\notag
\*B_0\mid\cdot \sim \sM\sN_{q \times d}\left(\left\{\sum_{i=1}^n\left( \widetilde{\boldsymbol{\varphi}}_i + \*B_0 \boldsymbol{x}_i \right)\boldsymbol{x}_i^\top\right\}\*S_{\omega}^{-1}, \*\Sigma, \*S_{\omega}^{-1}\right),
\end{equation}
where $\*S_\omega = \sum_{i=1}^n\boldsymbol{x}_i\boldsymbol{x}_i^\top + \*D_\omega$.

Similarly, for each $r \in [R]$, the full conditional of $\*B_r$ is
\begin{equation}
\notag
\*B_r\mid\cdot \sim \sM\sN_{q \times d}\left(\left\{\sum_{i=1}^n\gamma_{ir}\left( \widetilde{\boldsymbol{\varphi}}_i + \gamma_{ir}\*B_r \boldsymbol{x}_i \right)\boldsymbol{x}_i^\top\right\}\*S_{\nu,r}^{-1}, \*\Sigma, \*S_{\nu,r}^{-1}\right),
\end{equation}
where $\*S_{\nu,r} = \sum_{i=1}^n\gamma_{ir}^2\boldsymbol{x}_i\boldsymbol{x}_i^\top + \*D_\nu$.

For each $\gamma_{ir}$,
\begin{equation}
\notag
\gamma_{ir}\mid\cdot \sim \sN\left(s_{ir}^{-1} \left( \widetilde{\boldsymbol{\varphi}}_i +\gamma_{ir} \*B_r \boldsymbol{x}_i\right)^\top\*\Sigma^{-1}\*B_r\boldsymbol{x}_i, s_{ir}^{-1}\right),
\end{equation}
where $s_{ir} = 1+\boldsymbol{x}_i^\top\*B_r^\top\*\Sigma^{-1}\*B_r\boldsymbol{x}_i$.

For each $\boldsymbol{\varphi}_i$, we introduce Pólya-Gamma latent variables $g_{ik} \sim \text{PG}(N_{ik}, \varphi_{ik})$ for $k \in [q]$ \citep{polson2013bayesian}. Letting $\boldsymbol{\mu}_i = \*B_0\boldsymbol{x}_i+\sum_{r=1}^R\gamma_{ir}\*B_r\boldsymbol{x}_i$ and $\*O_i = \*\Sigma^{-1} + \text{diag}(\boldsymbol{g}_i)$, the full conditional of \(\boldsymbol{\varphi}_i\) is
\begin{equation}
\notag
\boldsymbol{\varphi}_i\mid\cdot \sim \sN_q\left(\*O_i^{-1}\Big(\boldsymbol{y}_i-\frac{\boldsymbol{N}_i}{2}+ \*\Sigma^{-1}\boldsymbol{\mu}_i\Big), \*O_i^{-1}\right).
\end{equation}

\subsection{Choice of hyperparameters}
Hyperparameters of our model include $v, s, a_\nu, b_\nu, a_\omega, b_\omega$, and the rank parameter $R$. The hyperparameters $v$ and $s$ determine the level of sparsity in $\*\Sigma$. We set $v = 1$ and $s = 0.01$ as recommended in \cite{wang2012bayesian}. The hyperparameters $(a_\nu, b_\nu)$ and  $(a_\omega, b_\omega)$ determine the amount of $L_2$ regularization imposed on $\set{\*B_r}_{r=1}^R$ and $\*B_0$, respectively, and they should be selected based on the relative size of the number of variables $q$ and the sample size $n$. Finally, the rank parameter $R$ determines the complexity of the model, and the true rank $R$ is unknown in real data applications. Given the hierarchical nature of our model, we select $R$ based on the Watanabe–Akaike information criterion (WAIC) \citep{watanabe2013widely}. WAIC is a Bayesian method for model selection, and is particularly suitable for complex hierarchical models such as ours. In numerical experiments, we found that WAIC leads to generally accurate rank selection.

\subsection{Computational complexity}
\label{subsec:comp_complexity}
The per-iteration cost of the Gibbs sampler is dominated by three operations: the augmented block update of $\*\Sigma$, which costs $O(q^3)$ via the graphical LASSO procedure of \cite{wang2012bayesian}; the matrix-normal draws for $\*B_0$ and $\set{\*B_r}_{r=1}^R$, each costing $O(q^2 d)$; and the multivariate-normal draws for $\set{\boldsymbol{\varphi}_i}_{i=1}^n$, each costing $O(q^2)$ due to $\*O_i^{-1}$. The overall per-iteration complexity is therefore $O(q^3 + nq^2 + Rq^2d)$.

Because the number of internal nodes $q$ in a binary phylogenetic tree equals the number of leaf taxa minus one, the matrix dimension in TRECOR is comparable to that of leaf-level methods such as MiCoRe \citep{mcgregor2020bayesian} for the same set of taxa. The per-iteration cost of the two methods therefore scales similarly with the network dimension. The computational advantage of the internal-node formulation lies instead in two aspects. First, the P\'{o}lya-Gamma augmentation of the binomial likelihood yields full conjugacy, so every parameter is updated by a Gibbs draw. By contrast, MiCoRe must resort to an adaptive Metropolis step for the latent additive log-ratio transformed proportions, because the multinomial likelihood is not conjugate to the Gaussian prior. Metropolis acceptance rates and mixing efficiency degrade rapidly in high dimensions, making the sampler increasingly impractical as the number of taxa grows. Second, internal-node counts aggregate over entire clades and are therefore far less zero-inflated than leaf-level counts, leading to better-conditioned likelihoods and faster convergence. These advantages compound as the network dimension grows, so the practical scalability of the proposed method is substantially better than what the per-iteration cost alone would suggest.

\section{Numerical experiments}
\label{sec:simu}

\subsection{Data generation}

We sampled synthetic data from the full hierarchical model \eqref{eq:expression_measurement_2} and  \eqref{eq:cov_reg_model} so that the observed taxa counts are compositional and the microbiome networks vary with covariates. 

We first generate covariates $\boldsymbol{x}_i \in \bR^d$ and latent Gaussian variables $\boldsymbol{\varphi}_i$ according to the covariance regression model \eqref{eq:cov_reg_model}. The first entry of $\boldsymbol{x}_i$ is an intercept. The remaining $d-1$ entries are sampled from a multivariate normal distribution with zero-mean, unit-variance, and pairwise correlation $0.1$. For the mean coefficient $\mathbf{B}_0$, we set all non-intercept columns to $0$ so that the mean abundance does not depend on the covariates. To mimic realistic abundance levels, the intercept column of $\mathbf{B}_0$ is estimated using data in \cite{yatsunenko2012human} (also analyzed in Section~\ref{sec:real_data}). The population covariance matrix $\*\Sigma$ is set to be sparse with unit diagonal. We consider three sparsity structures for $\*\Sigma$: tridiagonal, scale-free, and tree-based.  To generate the covariate-dependent covariance component, we fix the true rank $R$ and draw $\set{\*B_r}_{r=1}^R$ with non-overlapping supports across rows. Non-overlapping rows lead to mutually exclusive low-rank components for the covariate-dependent covariance structure. We draw the non-zero entries of $\*B_r$ as i.i.d. samples from a Gaussian distribution with mean zero and variance \(0.5d^{-1}\). 

Given latent Gaussian variables $\boldsymbol{\varphi}_i$, we generate observed counts using the tree-based expression-measurement model \eqref{eq:expression_measurement_2}. We generate a random binary phylogenetic tree with $q+1$ leaf nodes and map leaf node counts to internal-node binomial observations as \eqref{eq:expression_measurement_2}. The library sizes $N_{ij}$ are sampled to match the empirical distribution of data in \cite{yatsunenko2012human}. 

Across all experiments we fix $q = 100$, $d = 4$, and the true rank $R = 3$, and vary the sample size $n \in \set{150,250}$. This yields a total of $6$ settings including three $\*\Sigma$ structures and two sample sizes. Additional details including how we estimated mean abundance and library sizes from real data, and how each $\*\Sigma$ structure is generated, are provided in Section~\ref{subsec:par_gen} of the Appendix. 

\subsection{Performance evaluation} 

We evaluate performance in terms of both edge recovery and estimation/prediction accuracy:
\begin{itemize}
\item Edge selection for $\*\Sigma$ in terms of ROC curves for identifying nonzero off-diagonal entries of the population covariance $\*\Sigma$.
\item Posterior mean squared error for the estimation of $\*\Sigma$.
\item Prediction error for $\*\Sigma(\boldsymbol{x})$, defined as the posterior mean squared error 
\[
\frac{1}{n}\sum_{i=1}^n \left\|\widehat{\*\Sigma}(\boldsymbol{x}_i)-\*\Sigma(\boldsymbol{x}_i)\right\|_F^2 .
\]
\item Rank recovery: the proportion of replicates in which WAIC selects the true rank $R=3$.
\end{itemize}

We compare four approaches to highlight the roles of modeling the compositional sampling layer and incorporating covariates in network estimation.
\begin{enumerate}
    \item Sparse covariance estimation with a graphical-LASSO-type prior that ignores the covariates, fit to log-odds transformed binomial data (gLASSO).
    \item The proposed approach with the full model \eqref{eq:expression_measurement_2} and \eqref{eq:cov_reg_model} (TRECOR).
    \item The Gaussian covariance regression model \eqref{eq:cov_reg_model} fit to the latent Gaussian variables $\boldsymbol{\varphi}_i$, which serves as an oracle benchmark (TRECOR-oracle). The oracle uses the same model but bypasses the compositional sampling layer entirely. Any method operating on observed count data must incur additional estimation error from recovering the latent variables and therefore cannot exceed oracle performance under the same model. The gap between TRECOR and TRECOR-oracle thus quantifies the information loss attributable to the compositional sampling layer, and is expected to narrow with larger sample sizes or less sparse data.
    \item The Gaussian covariance regression model \eqref{eq:cov_reg_model} fit to log-odds transformed binomial data (CovReg).
\end{enumerate}
For all four methods, we run MCMC chains of 10{,}000 iterations and discard the first 5{,}000 as burn-in. For the three covariance regression methods, we run with rank $R \in \set{1,2,3,4,5}$ and select the optimal rank by WAIC. Unless otherwise noted, we use the hyperparameters $v = 1, s = 0.01, a_\nu = 5, b_\nu = 0.5, a_\omega = 5$ and $b_\omega = 0.5$. The hyperparameters $a_\nu$, $b_\nu$, $a_\omega$ and $b_\omega$ determine the strength of $L_2$ penalties on the high-dimensional parameters $\*B_0$ and $\set{\*B_r}_{r=1}^R$ and are specified based on the relative size of $q$ and $n$. Sensitivity of the proposed method on these hyperparameters is presented in Section~\ref{subsec:apdx_simu_results} of the Appendix.

We also note that direct comparison of our approach with leaf-node covariance regression methods such as MiCoRe \citep{mcgregor2020bayesian} is not straightforward, as these methods estimate covariation in a fundamentally different representation space. The transformation from internal-node covariance to leaf-node covariance is nonlinear \citep{greenacre2020amalgamations}. \cite{wang2021tree} and \cite{hong2023phylomed} have noted the same challenge in the context of differential abundance and mediation analysis on internal nodes. We therefore focus the comparison on methods that operate in the same internal-node space.

\subsection{Results}

\begin{figure}[h]
    \centering
    \includegraphics[width=0.95\linewidth]{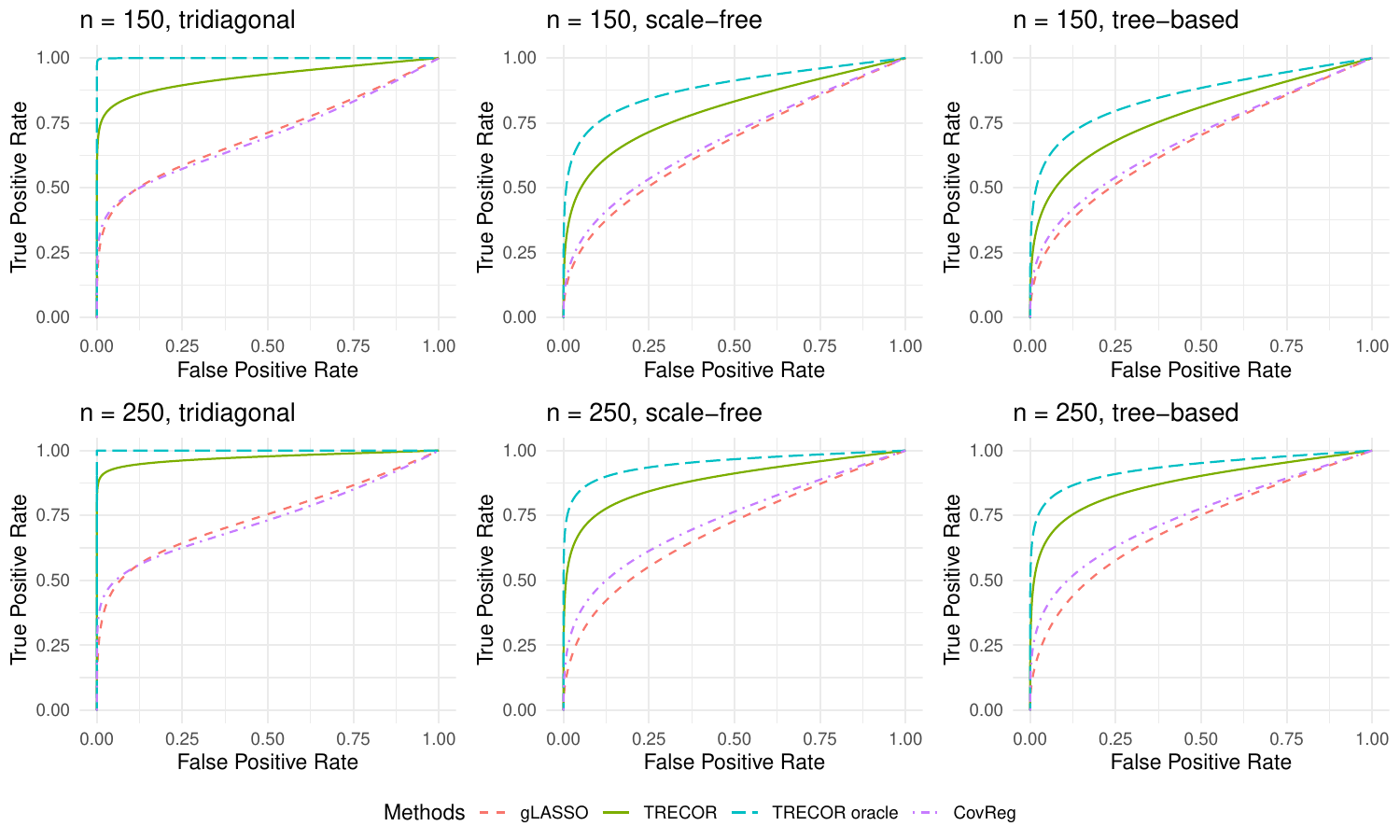}
    \caption{Average ROC curves for recovering nonzero off-diagonal entries of the population covariance matrix $\*\Sigma$ across 100 replications. Panels correspond to $n\in\{150,250\}$ and three sparse structures for $\*\Sigma$ (tridiagonal, scale-free, and tree-based). Curves compare gLASSO (coral dashed), TRECOR (green solid), TRECOR-oracle (light blue longdash) and CovReg (violet dotdash).}
    \label{fig:simu_ROC}
\end{figure}

Figure~\ref{fig:simu_ROC} and Table~\ref{tab:simu_mnse} report the average ROC curve and accuracy measures over 100 replications, respectively. Across all settings, the proposed method achieves the second-best performance next to the oracle benchmark, which uses the unobserved latent variables. In contrast, fitting the Gaussian covariance regression model to log-odds transformed data substantially degrades both edge recovery and estimation accuracy, underscoring the importance of explicitly modeling the compositional sampling layer rather than relying on transformations to approximate Gaussianity. Finally, naively performing graphical-LASSO leads to the worst performance, particularly for predicting $\*\Sigma(\boldsymbol{x})$, because it does not incorporate covariate effects. Overall, these results indicate that accounting for compositionality and modeling covariate-dependent covariance are essential for constructing microbiome networks that vary with host or environmental factors. Additional simulation results, including the precision-recall curves, can be found in Section~\ref{subsec:apdx_simu_results} of the Appendix.

\begin{table}[ht]
\centering
\centering
\begin{threeparttable}
\caption{Average estimation and prediction accuracy over 100 replications (with standard deviation in parentheses). ``Error of $\*\Sigma$'' is the posterior mean squared error for the estimation of $\*\Sigma$. ``Error of $\*\Sigma(\boldsymbol{x})$'' is the prediction error for the estimation of $\*\Sigma(\boldsymbol{x})$. ``Rank selection'' reports the proportion of replicates in which WAIC selects the true rank $R=3$; NA indicates the criterion is not applicable. }
\label{tab:simu_mnse}
\small
\setlength{\tabcolsep}{5pt}
\renewcommand{\arraystretch}{1.1}
\begin{tabular}{|c|c| l r r r|}
  \hline
 n & Structure & Method & Error of $\*\Sigma$ &  Error of $\*\Sigma(\boldsymbol{x})$ & Rank selection\\ 
 \hline
 \multirow{12}{*}{150} & \multirow{4}{*}{tridiagonal} & gLASSO & 6429.67 (532.47) & 27421.66 (1116.39) & NA \\
  & & TRECOR & 33.82 (1.59) & 3489.09 (430.77) & 0.80  \\
  & & TRECOR-oracle & 19.83  (0.44) & 2539.96 (235.16) & 1.00 \\
  & & CovReg & 2769.26 (228.07) & 70091.72 (7305.81) & 0.09  \\
  \cline{2-6}

  & \multirow{4}{*}{scale-free} & gLASSO & 6511.64  (515.84) & 27343.50 (1165.56) & NA \\
  & & TRECOR & 35.38 (1.51) & 3527.59 (431.08) & 0.79 \\
  & & TRECOR-oracle & 28.50 (0.76) & 2449.96 (289.84) & 0.98 \\
  & & CovReg & 2800.14 (216.05) & 68827.76 (5772.26) & 0.09 \\
  \cline{2-6}

  & \multirow{4}{*}{tree-based} & gLASSO & 6564.67  (559.12) & 27413.01 (1236.63) & NA \\
  & & TRECOR & 35.70 (1.60) & 3538.42 (416.22) & 0.81\\
  & & TRECOR-oracle & 29.56 (0.82) & 2534.08  (293.31) & 0.97\\
  & & CovReg & 2861.09 (271.01) &  67486.99 (7570.74) & 0.02\\
  \hline

  \multirow{12}{*}{250} & \multirow{4}{*}{tridiagonal} & gLASSO & 7360.11 (443.14) & 29365.82 (911.83) & NA \\
  & & TRECOR & 28.24 (0.96)  & 2105.52 (329.21) & 0.76 \\
  & & TRECOR-oracle & 13.68 (0.23) & 1695.73 (201.59) & 1.00 \\
  & & CovReg & 3426.68 (228.36) & 55970.57 (4899.53) & 0.01 \\
  \cline{2-6}

  & \multirow{4}{*}{scale-free} & gLASSO & 7354.44 (393.55) & 29123.14 (876.85) & NA \\
  & & TRECOR &  29.59 (1.19) & 2174.10 (378.42) & 0.79 \\
  & & TRECOR-oracle & 20.33 (0.48) & 1310.51 (190.62) & 0.96 \\
  & & CovReg & 3430.78 (236.89) & 54078.07 (4541.57) & 0.07 \\
  \cline{2-6}

  & \multirow{4}{*}{tree-based} & gLASSO & 7412.50 (413.93) & 29286.88 (928.21) & NA \\
  & & TRECOR & 29.81 (1.09) & 2180.49 (362.28) & 0.79\\
  & & TRECOR-oracle & 21.42  (0.49) & 1490.78  (221.32) & 0.97\\
  & & CovReg & 3479.71 (262.51) & 53857.64 (4739.28) & 0.02 \\
   \hline
\end{tabular}
\end{threeparttable}
\end{table}

To put the complexity analysis of Section~\ref{subsec:comp_complexity} in concrete terms, we report wall-clock times for the simulation and compare with MiCoRe. On local computing clusters with Intel(R) Xeon(R) Gold 6254 CPU at 3.10GHz processor with 376 GB of RAM, running 10{,}000 MCMC iterations for TRECOR with $q = 100, n = 150, d=3$ and $R=3$ takes approximately 2.6 hours per chain. 
For smaller networks (e.g., $q = 25$ as in \cite{mcgregor2020bayesian}), the average running time of TRECOR is 51 minutes, which is similar to that of MiCoRe and consistent with both methods scaling as $O(q^3)$ in the network dimension. The key difference is that TRECOR's fully conjugate sampler scales reliably to $q = 100$ internal nodes without the convergence difficulties that would arise from running MiCoRe's Metropolis-within-Gibbs sampler. Indeed, MiCoRe's simulations were limited to $p \leq 25$ and required filtering of low-count samples to ensure convergence.

\section{Case study}
\label{sec:real_data}

\subsection{Data description and preprocessing}
We analyzed a gut microbiome data set from \cite{yatsunenko2012human}, consisting of fecal samples profiled by 16S rRNA sequencing from 531 individuals. This cohort includes both children and adults from three geographically and culturally distinct regions: Venezuela, Malawi and the United States. While \cite{yatsunenko2012human} demonstrated that there exists substantial variation in microbial abundances across age groups and geographical locations, how these factors influence microbial co-variation remains largely unexplored. {Prior to analysis, we filtered out rare genera with zero proportion exceeding $0.5$ or total counts less than $100$.} Using the phylogenetic tree, we converted genus-level counts to counts associated with internal tree nodes. Our model included an intercept and three categorical covariates: sex (binary), age (binary indicator of whether an individual was older than 3 years), and country, with Malawi being the reference group. Subjects with missing covariate information were excluded. After pre-processing, the final data set comprised $n = 476$ samples, 90 genera or equivalently 89 internal nodes, and $d = 5$ covariates, corresponding to a scenario with high-dimensional responses and low-dimensional covariates. 

We used the same hyperparameters and MCMC settings as in the simulation studies (Section~\ref{sec:simu}). We selected the rank from $R \in \set{1,2,3,4,5}$ by WAIC, choosing $R = 4$ which corresponds to the elbow of the WAIC curve.

\subsection{Results}
\begin{figure}[h]
    \centering
    \includegraphics[width=\textwidth]{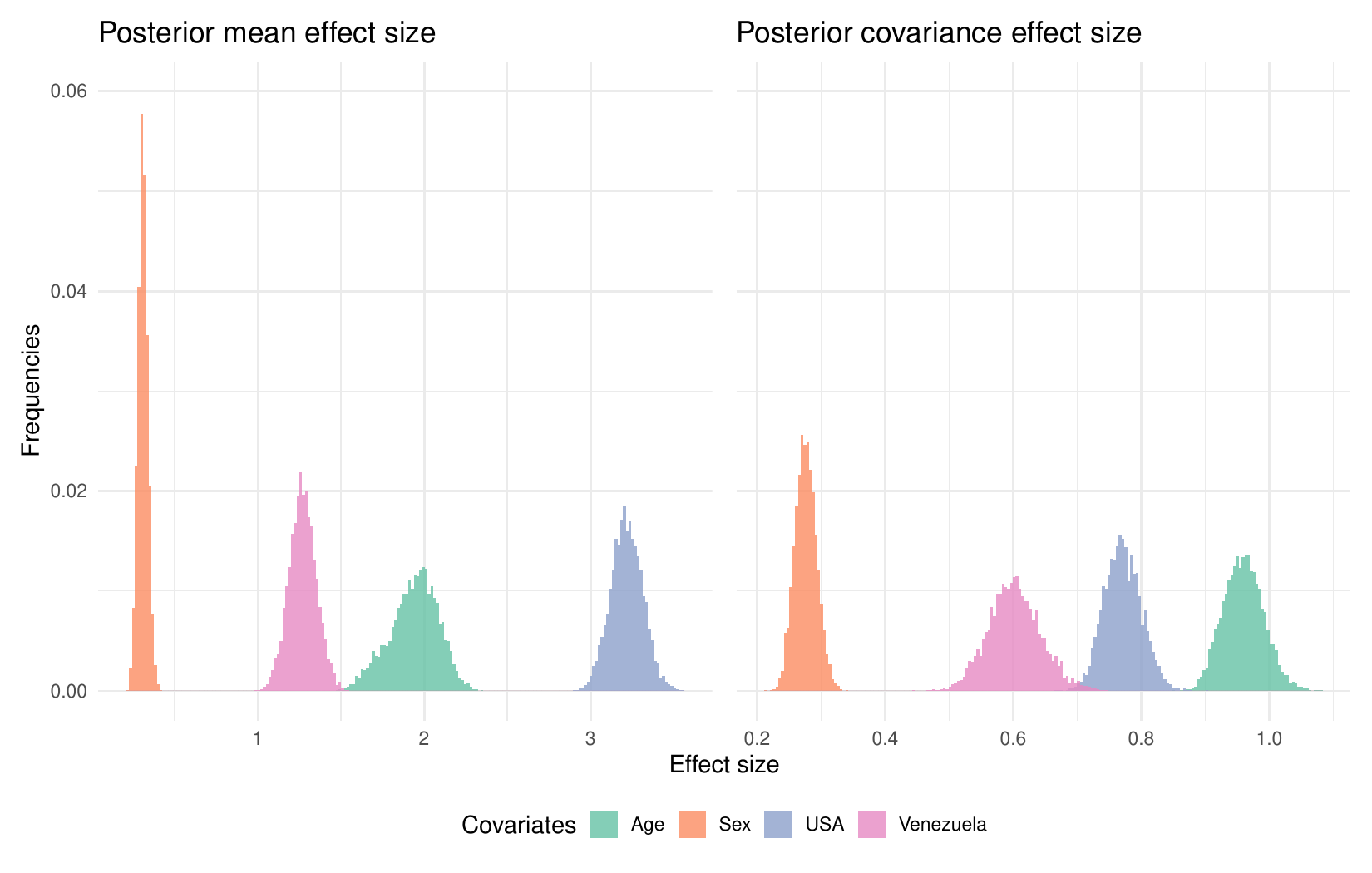} 
    \caption{Left panel: posterior mean effect size distribution for each covariate (excluding intercept), represented by the posterior distribution of $\norm{\boldsymbol{b}_{j0}}_2^2$ where $\boldsymbol{b}_{j0} \in \bR^{q}$ is the $j$th column of $\*B_0$. Right panel: posterior covariance effect size distribution for each covariate (excluding intercept), represented by the posterior distribution of $\norm{\*B^{(j)}}_F$.}
    \label{fig:post_effect_size_hist}
\end{figure}

We first examined the effects of covariates on the \textit{mean structure of taxonomic abundances}. To this end, we compared the posterior effect size distribution of each covariate by sampling from the posterior of the Euclidean norm of the corresponding columns of $\mathbf{B}_0$. The left panel of Figure~\ref{fig:post_effect_size_hist} displays the posterior effect size distributions for all four covariates (excluding intercept). It is clear that sex has the smallest effect on the mean, with posterior effect size distribution concentrating around $0$. Using Malawi and age less than 3 years as reference categories, the indicator for the United States shows the largest posterior mean effect, followed by age, whereas the effect associated with Venezuela is substantially smaller. These findings suggest that average microbial abundances differ substantially between individuals younger than 3 years and those aged 3 years or older, and between U.S. and non-U.S. populations, consistent with the results of  \cite{yatsunenko2012human}.

Next, we examined effects of covariates on the \textit{covariance structure of microbial abundances}. Posterior effect size distributions were obtained by the posterior distribution of $\|{\mathbf{B}^{(j)}}\|_{F}$ for each $j \in [d]$. Note that the Frobenius norm of the coefficient matrix is invariant to orthogonal rotations and is therefore identifiable according to Lemma~\ref{lem:identifiablity}. The right panel of Figure~\ref{fig:post_effect_size_hist} shows the posterior distribution of $\|{\mathbf{B}^{(j)}}\|_{F}$ for all four covariates (excluding intercept). As in the analysis of mean abundances, sex has the smallest effect sizes among all covariates. Relative to Malawi, the effect size distribution of the indicator for the United States is larger in magnitude compared to that for Venezuela. However, in contrast to the mean abundance analysis, age has the largest effect on covariance. This highlights that covariance regression analysis can uncover unique patterns not revealed by mean-based analysis.  

We further investigated the effects of age and country on microbial covariance by constructing the corresponding differential correlation matrix. We did not consider sex, as its effect was much smaller compared to age and country. Let $\boldsymbol{x}_{0} = [1,0,0,0,0]^\top$ represent the baseline covariate. For the $j$-th covariate, consider the vector $\boldsymbol{x}_{j}$ which equals $\boldsymbol{x}_{0}$ except that its $j$-th entry is $1$. We obtained the posterior distribution of the differences between correlation matrices evaluated at $\boldsymbol{x}_{0}$ and $\boldsymbol{x}_j$, and scaled the posterior distribution by $1/2$ to keep the entries within the range $[-1,1]$. We then computed the posterior mean of this distribution and employed an additional filtering step to filter out insignificant off-diagonal entries. The resulting matrix is treated as the differential correlation matrix relative to the $j$-th covariate. Details of constructing the differential correlation matrix and the filtering step are described in Section~\ref{subsec:apdx_diff} of the Appendix. The differential correlation matrices corresponding to age, the indicator of USA, and the indicator of Venezuela consist of 258, 201, and 12 non-zero off-diagonal entries, respectively. On each differential network, we computed the degree of each internal node and identified the node with the highest degree, i.e., the node with the most significant differential correlations. The genera corresponding to the leaf descendants of this node form what we refer to as the differential set, and represent the taxa whose collective abundance patterns are most strongly affected by the covariate. 
We found that the differential set with respect to age is highly enriched for taxa in \textit{Enterobacteriaceae} and related families. These taxa include facultative anaerobes that are prominent in the neonatal and early-infant gut and tend to decrease over infancy as the microbiota matures and obligate anaerobes become more dominant \citep{robertson2019human, sanidad2020neonatal}.  The differential sets corresponding to the indicators of the US and Venezuela both consist of diet-related taxa, highlighting the role of dietary differences in shaping microbiome co-variation relative to Malawi. However, the differential network corresponding to the US has many more significant edges and much larger non-zero entries compared to that of Venezuela. This is consistent with \cite{yatsunenko2012human}, as dietary patterns of Venezuela are more similar to those in Malawi than are those of the United States.

\begin{figure}[htbp]
  \centering
  \includegraphics[width=0.5\textwidth]{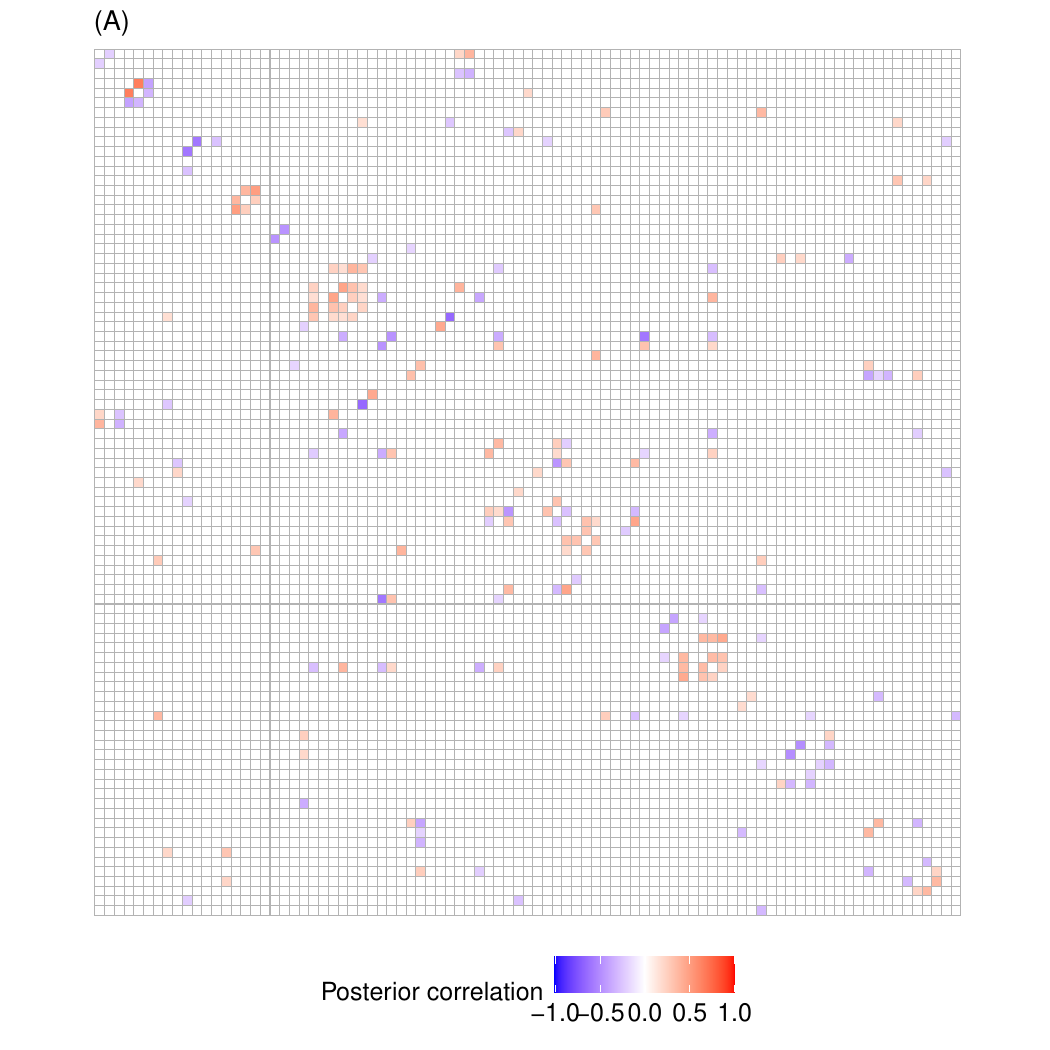}\hfill
  \includegraphics[width=0.5\textwidth]{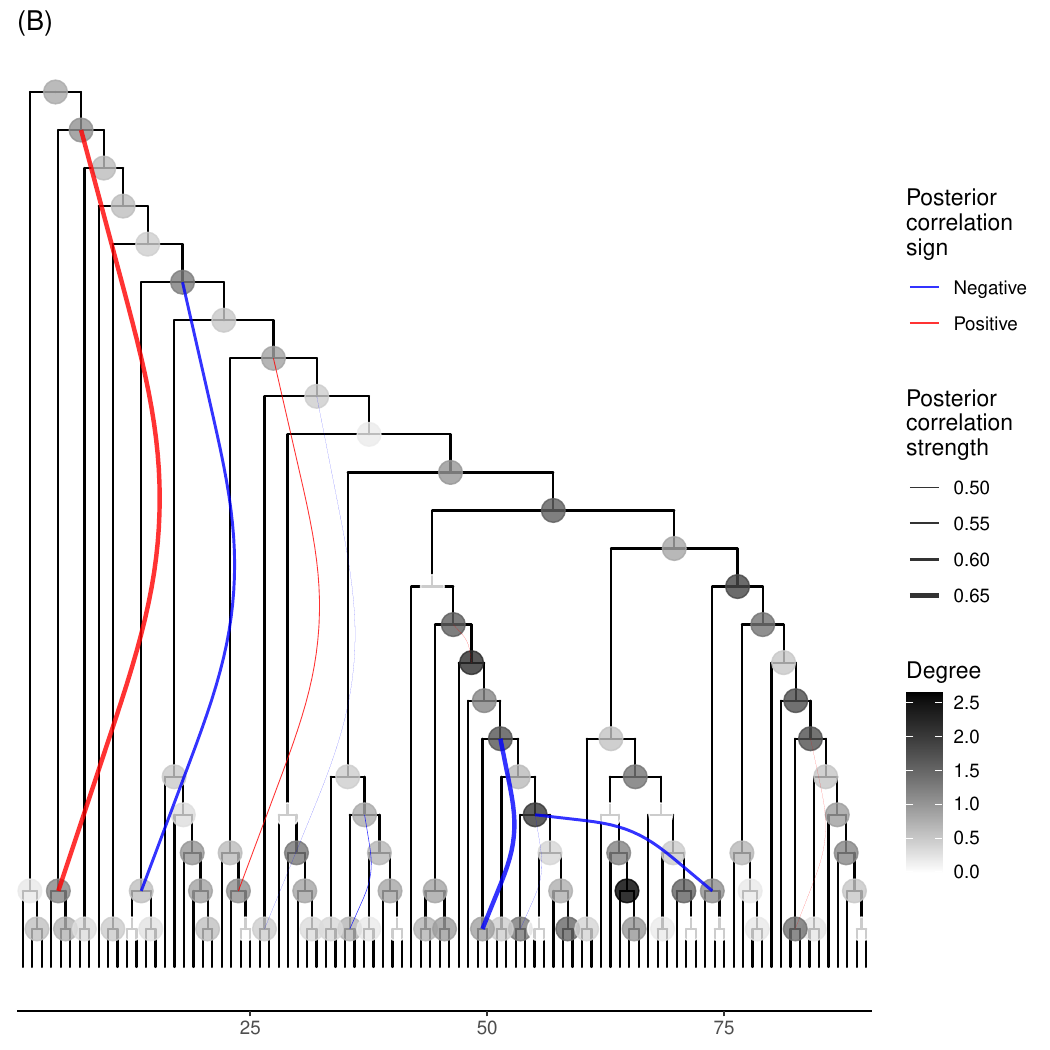}
  \caption{(A) Posterior mean of the  population correlation matrix $\*\Sigma$, after filtering out statistically insignificant off-diagonal entries. Rows and columns are ordered according to a depth-first traversal of the phylogenetic tree. (B) The 10 largest correlations in absolute value from the population correlation matrix, overlaid on the phylogenetic tree. Red and blue edges indicate positive and negative network links, respectively, with line width proportional to the absolute posterior mean correlation. Node color indicates the degree of each node in the correlation network.}
  \label{fig:population_network}
\end{figure}

Finally, we estimated the population correlation matrix using the posterior mean of $\*\Sigma$ with the same filtering procedure to remove insignificant edges. The resulting matrix contains 111 nonzero off-diagonal entries, representing baseline correlations among internal nodes after removing covariate-dependent effects. The left panel of Figure~\ref{fig:population_network} displays this population correlation matrix, with rows and columns ordered according to a depth-first traversal of the phylogenetic tree, so that adjacent variables in the heatmap also tend to be close on the tree. The matrix is sparse overall, but nonzero entries cluster near the diagonal, indicating that significant baseline correlations occur primarily between nodes that are close in the tree topology. {Such a structure is much less apparent in the differential correlation matrices}. To further examine this structure, the right panel of Figure~\ref{fig:population_network} overlays the 10 largest correlations in absolute value on the tree. All of the top correlations are between parent--child node pairs: the log-ratio of the branching probability at a parent node is naturally correlated with those at its children, since the children govern finer partitions within the same group of taxa. Among the few negative correlations, the log-odds at a child node moves in the opposite direction from that of its parent, suggesting compositional trade-offs within a clade. A further notable pattern is that nodes with the highest degrees tend to be leaf-adjacent, meaning that a small number of fine-level log-odds act as hubs, each participating in multiple strong parent--child correlations.

\subsection{Sensitivity analysis and MCMC convergence}
To assess MCMC convergence, we examined trace plots of randomly selected parameters and ran three independent chains with different initializations; all diagnostics indicated good convergence. To validate the robustness of our findings, we performed sensitivity analysis over rank parameters $R \in \set{3,4,5}$; results were insensitive to the choice of $R$. Details are given in Section~\ref{subsec:apdx_convergence} of the Appendix.

\section{Discussion}

We have proposed TRECOR, a tree-based covariance regression framework for differential network analysis of zero-inflated microbiome data. A key design choice in the proposed method is to model covariation at the level of internal nodes of a phylogenetic tree rather than at the level of individual taxa. This choice confers several statistical advantages. First, internal-node counts aggregate over entire clades, which substantially mitigates the zero inflation that plagues leaf-level count data and leads to more stable covariance estimates. Second, by exploiting the binomial factorization of the multinomial distribution at each internal node, the model achieves full conjugacy, enabling efficient Gibbs sampling that scales to high-dimensional settings. Third, each internal node has a direct phylogenetic interpretation as a clade-level log ratio, providing a natural hierarchical summary of the microbial community that is more interpretable than taxa at a fixed rank.

Beyond these statistical advantages, the internal-node representation opens a natural connection to ecological concepts of community structure. A central question when analyzing covariation networks is which nodes play the most central role in structuring the community. In the ecology literature, this concept is captured by the notion of a keystone species: a taxon whose presence or removal disproportionately affects the rest of the community \citep{paine1995conversation,agler2016microbial}. In our framework, internal nodes with high degrees in the estimated covariation network, i.e., those whose branching log-odds co-vary strongly with many other internal nodes, may represent keystone clades, whose abundance patterns are tightly coupled with the broader community structure. In our data analysis, we identified such high-degree nodes in both the population network and the differential networks, and the leaf taxa descending from these nodes were enriched for biologically meaningful groups. This suggests that degree-based screening of the internal-node covariation network may offer a tractable approach to keystone clade identification.

Despite these strengths, the internal-node representation also introduces an important limitation. The connection between internal-node covariation and leaf-level taxon covariation is not straightforward. The internal-node log-odds are nonlinearly related to the original taxon relative abundances \citep{greenacre2020amalgamations}, and no closed-form transformation exists from internal-node covariance to leaf-level correlation. As a practical approximation, following \cite{hong2023phylomed}, we identify the leaf descendants of high-degree internal nodes as differentially co-varying taxa. While this provides interpretable biological summaries, developing a principled framework for translating internal-node covariance findings to leaf-level statements remains an open methodological challenge and an important direction for future work.

\section{Data and Code Availability}
All data and code used to reproduce the results in this paper are available at \url{https://github.com/zichun-xu/TRECOR}.

\section{Funding}

JM is supported by NIH R01 GM145772 and R01 GM151301.

\bibliography{ref}

\appendix

\section{Details of the Gibbs algorithm}
\label{sec:apdx_gibbs}
We provide derivations of the full conditional distributions used in the Gibbs algorithm of Section~2.4 of the main text.

Let $f_{\sN_q}(\cdot \mid \boldsymbol{\mu}, \*\Sigma)$ denote the density of the $q$-dimensional normal distribution with mean $\boldsymbol{\mu}$ and covariance $\*\Sigma$. The full conditional of $\*\Sigma$ is
\begin{equation*}
\begin{aligned}
\pi(\*\Sigma, \boldsymbol{\tau}\mid \cdot) &\propto \pi(\*\Sigma\mid \boldsymbol{\tau},  \lambda)\pi(\boldsymbol{\tau}\mid\lambda)\prod_{i=1}^n\left\{ f_{\sN_q}\left(\boldsymbol{\varphi}_i \mid \*B_0\boldsymbol{x}_i+\sum_{r=1}^R\gamma_{ir}\*B_r\boldsymbol{x}_i,  \*\Sigma\right)\right\} \\
&\qquad  \times \pi(\*B_0\mid \*\Sigma, \*D_\omega) \prod_{r=1}^R\pi(\*B_r\mid \*\Sigma, \*D_\nu) \\
&\propto \pi(\*\Sigma, \boldsymbol{\tau}\mid  \lambda)\det(\*\Sigma)^{-\frac{n+(R+1)d}{2}}\exp\set{-\frac{1}{2}\tr(\*S\*\Sigma^{-1})},
\end{aligned}
\end{equation*}
where $\*S$ is the sufficient statistic defined in the main text. This structure enables the augmented block Gibbs update of \cite{wang2012bayesian}.

The full conditional of $\*B_0$ is
\begin{equation}
\notag
\begin{aligned}
\pi(\*B_0\mid \cdot) &\propto \pi(\*B_0\mid \*\Sigma, \*D_\omega)\prod_{i=1}^n\left\{f_{\sN_q}\left(\boldsymbol{\varphi}_i - \sum_{r=1}^R\gamma_{ir}\*B_r \boldsymbol{x}_i\mid \*B_0\boldsymbol{x}_i, \*\Sigma\right)\right\} \\
&\sim \sM\sN_{q \times d}\left(\left\{\sum_{i=1}^n\left(\boldsymbol{\varphi}_i-\sum_{r=1}^R\gamma_{ir}\*B_r\boldsymbol{x}_i\right)\boldsymbol{x}_i^\top\right\}\*S_{\omega}^{-1}, \*\Sigma, \*S_{\omega}^{-1}\right).
\end{aligned}
\end{equation}
Similarly, for each $r \in [R]$, the full conditional of $\*B_r$ is
\begin{equation}
\notag
\begin{aligned}
\pi(\*B_r\mid \cdot) &\propto \pi(\*B_r\mid \*\Sigma, \*D_\nu)\prod_{i=1}^n\left\{f_{\sN_q}\left(\boldsymbol{\varphi}_i - \*B_0 \boldsymbol{x}_i - \sum_{r^\prime \ne r}^R\gamma_{ir^\prime}\*B_{r^\prime}\boldsymbol{x}_i\mid \gamma_{ir}\*B_r\boldsymbol{x}_i, \*\Sigma\right)\right\} \\
&\sim \sM\sN_{q \times d}\left(\left\{\sum_{i=1}^n\gamma_{ir}\left(\boldsymbol{\varphi}_i-\*B_0\boldsymbol{x}_i - \sum_{r^\prime \ne r}^R\gamma_{ir^\prime}\*B_{r^\prime}\boldsymbol{x}_i\right)\boldsymbol{x}_i^\top\right\}\*S_{\nu,r}^{-1}, \*\Sigma, \*S_{\nu,r}^{-1}\right).
\end{aligned}
\end{equation}

For each $\gamma_{ir}$,
\begin{equation}
\notag
\begin{aligned}
\pi(\gamma_{ir} \mid \dot) &\propto f_\sN(\gamma_{ir} \mid 0, 1) f_{\sN_q}\left(\boldsymbol{\varphi}_i - \*B_0\boldsymbol{x}_i - \sum_{r^\prime\ne r}^R\gamma_{ir^\prime}\*B_{r^\prime}\boldsymbol{x}_i\mid \gamma_{ir}\*B_r \boldsymbol{x}_i, \*\Sigma \right) \\
&\sim \sN\left(s_{ir}^{-1}\left( \boldsymbol{\varphi}_i - \*B_0\boldsymbol{x}_i - \sum_{r^\prime\ne r}^R\gamma_{ir^\prime}\*B_{r^\prime}\boldsymbol{x}_i\right)^\top\*\Sigma^{-1}\*B_r\boldsymbol{x}_i, s_{ir}^{-1}\right).
\end{aligned}
\end{equation}

The full conditional of $\nu_j$ for $j \in [d]$ can be written as:
\begin{equation}
\notag
\begin{aligned}
\pi(\nu_j\mid \cdot) &\propto \pi(\nu_j) \prod_{r=1}^R\pi(\boldsymbol{b}_{jr}\mid \*\Sigma, \nu_j)\sim \text{IG}\left(a_\nu+\frac{qR}{2}, b_\nu + \frac{1}{2}\sum_{r=1}^R\boldsymbol{b}_{jr}^\top\*\Sigma^{-1}\boldsymbol{b}_{jr}\right).
\end{aligned}
\end{equation}
Similarly, the full conditional of $\omega_j$ for $j \in [d]$ is:
\begin{equation}
\notag
\begin{aligned}
\pi(\omega_j\mid \cdot) &\propto \pi(\omega_j) \pi(\boldsymbol{b}_{j0}\mid \*\Sigma, \omega_j)\sim \text{IG}\left(a_\omega+\frac{q}{2}, b_\omega + \frac{1}{2}\boldsymbol{b}_{j0}^\top\*\Sigma^{-1}\boldsymbol{b}_{j0}\right).
\end{aligned}
\end{equation}

The full Gibbs algorithm is summarized as follows.
\begin{algorithmic}[1]
\Require Observed count data  $\set{\boldsymbol{z}_i}_{i=1}^n$, binary phylogenetic tree $\sT$, covariates $\set{\boldsymbol{x}_i}_{i=1}^n$, hyperparameters $v, s, a_\nu, b_\nu, a_\omega, b_\omega$, and rank $R$, initial values $\*\Sigma^{(0)}$, $\set{\*B_r^{(0)}}_{r=0}^R$, $\boldsymbol{\tau}^{(0)}$, $\lambda^{(0)}$, $\set{\nu_j^{(0)}}_{j=1}^d$, $\set{\omega_j^{(0)}}_{j=1}^d$, $\set{\boldsymbol{\varphi}^{(0)}_i}_{i=1}^n$, $\set{\boldsymbol{g}^{(0)}_i}_{i=1}^n$, and $\set{\boldsymbol{\gamma}^{(0)}_i}_{i=1}^n$, and iterations $T$.
\State Convert the count data $\set{\boldsymbol{z}_i}_{i=1}^n$ to internal-node level $\set{(\boldsymbol{N}_i, \boldsymbol{y}_i)}_{i=1}^n$.
\For{$t \in [T]$}
    \For{$i \in [n]$}
        \State Update
        \[
         \widetilde{\boldsymbol{\varphi}}^{(t)}_i \leftarrow \boldsymbol{\varphi}^{(t-1)}_i - \*B_0^{(t-1)}\boldsymbol{x}_i - \sum_{r=1}^R\gamma_{ir}^{(t-1)}\*B_r^{(t-1)}\boldsymbol{x}_i;
         \]
    \EndFor
    \State Update
    $$
    \*S^{(t)} \leftarrow \sum_{i=1}^n\widetilde{\boldsymbol{\varphi}}^{(t)}_i\left(\widetilde{\boldsymbol{\varphi}}^{(t)}_i\right)^\top + \sum_{r=1}^R\*B_r^{(t-1)}\left(\*D^{(t-1)}_\nu\right)^{-1}\left(\*B^{(t-1)}_r\right)^\top + \*B^{(t-1)}_0\left(\*D^{(t-1)}_\omega\right)^{-1}\left(\*B^{(t-1)}_0\right)^\top;
    $$
    \For{$k \in [q]$}
        \State Update
        \[
        \boldsymbol{H}^{(t)}\leftarrow\left(\*\Sigma_{-k,-k}^{(t-1)}\right)^{-1};
        \]
        \State Update
        \begin{align*}
        u_k^{(t-1)} & \leftarrow \*\Sigma_{k,k}^{(t-1)}-\left(\*\Sigma^{(t-1)}_{k,-k}\right)^\top\boldsymbol{H}^{(t)}\*\Sigma^{(t-1)}_{k,-k},\\
        \*\Theta_k^{(t)} & \leftarrow \left(u_k^{(t-1)}\right)^{-1}\boldsymbol{H}^{(t)}\*S_{-k,-k}^{(t)}\boldsymbol{H}^{(t)} +\lambda^{(t-1)} \boldsymbol{H}^{(t)}+\text{diag}\left(\boldsymbol{\tau}_{k,-k}^{(t-1)}\right),
        \end{align*}
        where $\boldsymbol{\tau}^{t-1}_{k,-k} = \set{\tau^{(t-1)}_{k,l}}_{l \ne k}$;
        \State Sample
        $$\*\Sigma_{k,-k}^{(t)} \leftarrow \sN_{q-1}\left(\left[u_k^{(t-1)}\right]^{-1}\left[\*\Theta_k^{(t)}\right]^{-1}\boldsymbol{H}^{(t)}\*S^{(t)}_{k,-k}, \left[\*\Theta_k^{(t)}\right]^{-1}\right);
        $$
        
        \State Update
        \begin{equation}
        \notag
        \begin{aligned}
        \chi^{(t)} \leftarrow \left(\*\Sigma^{(t)}_{k,-k}\right)^\top\boldsymbol{H}^{(t)}\*S^{(t)}_{-k,-k}\boldsymbol{H}^{(t)}\*\Sigma^{(t)}_{k,-k} - 2\left(\*S_{k,-k}^{(t)}\right)^\top \boldsymbol{H}^{(t)}\*\Sigma^{(t)}_{k,-k} + \*S^{(t)}_{k,k};
        \end{aligned}
        \end{equation}
        \State Sample $u_k^{(t)} \sim \text{GIG}\left(1-\frac{n+d(R+1)}{2}, \lambda^{(t-1)}, \chi^{(t)}\right),$ where GIG refers to the generalized inverse Gaussian distribution;
        \State Update $\*\Sigma^{(t)}_{k,k} \leftarrow u_k^{(t)} + \left(\*\Sigma^{(t)}_{k,-k}\right)^\top\boldsymbol{H}^{(t)}\*\Sigma^{(t)}_{k,-k}$;
    \EndFor
    \For{$k \ne l \in [q]$}
    	\State sample
    \[
    \left(\tau^{(t)}_{k,l}\right)^{-1} \sim \text{IG}\left(1,\lambda^{(t-1)}\left\lvert \*\Sigma^{(t)}_{k,l} \right\rvert^{-1}, \left(\lambda^{(t-1)}\right)^2\right);
    \]
    \EndFor
    \State Sample
    \[
    \lambda^{(t)} \sim \text{Gamma}\left(1, v+\frac{q(q-1)}{2}, s+\frac{\norm{\*\Sigma^{(t)}}_1}{2}\right);
    \]
    \For{$r \in [R], i \in [n]$}
        \State Update
        \[
        s_{ir}^{(t)} \leftarrow 1+\boldsymbol{x}_i^\top\left(\*B_r^{(t-1)}\right)^\top\left(\*\Sigma^{(t)}\right)^{-1}\*B_r^{(t-1)}\boldsymbol{x}_i;
        \]
        \State Update
        \[
        \mathbf{V}_{ir}^{(t)} \leftarrow\boldsymbol{\varphi}_i^{(t-1)} - \*B_0^{(t-1)}\boldsymbol{x}_i - \sum_{r^\prime\ne r}^R\gamma_{ir^\prime}^{(t-1)}\*B_{r^\prime}^{(t-1)}\boldsymbol{x}_i;
        \]
        \State Sample $$\gamma_{ir}^{(t)} \sim \sN\left(\left(s_{ir}^{(t)}\right)^{-1}\Big(\mathbf{V}_{ir}^{(t)}\Big)^\top\left(\*\Sigma^{(t)}\right)^{-1}\*B_r^{(t-1)}\boldsymbol{x}_i, \left(s_{ir}^{(t)}\right)^{-1}\right);
        $$
    \EndFor
    \For{$r \in [R]$}
    	\State Update
	$$
	\*S_{\nu,r}^{(t)} \leftarrow \sum_{i=1}^n\left(\gamma_{ir}^{(t)}\right)^2\boldsymbol{x}_i\boldsymbol{x}_i^\top + \*D_\nu^{(t-1)};
	$$
    
    \State    Update 
    \[
    \mathbf{W}_{ir}^{(t)} \leftarrow \boldsymbol{\varphi}^{(t-1)}_i - \mathbf{B}_0^{(t-1)}\boldsymbol{x}_i - \sum_{r^\prime \ne r}^R \gamma_{ir^\prime}^{(t)} \mathbf{B}_{r^\prime}^{(t-1)}\boldsymbol{x}_i  .
    \]
        \State Sample $$\*B_r^{(t)} \sim \sM\sN_{q \times d}\left(\Bigg(\sum_{i=1}^n\gamma_{ir}^{(t)} \mathbf{W}_{ir}^{(t)} \boldsymbol{x}_i^\top\Bigg)\left(\*S_{\nu,r}^{(t)}\right)^{-1}, \*\Sigma^{(t)}, \left(\*S_{\nu,r}^{(t)}\right)^{-1}\right);
        $$
    \EndFor
    \State Update
    \[
    \*S_{\omega}^{(t)} \leftarrow \sum_{i=1}^n\boldsymbol{x}_i\boldsymbol{x}_i^\top + \*D_\omega^{(t-1)};
    \]
    \State Sample
    $$\*B_0^{(t)} \sim \sM\sN_{q \times d}\left(\left\{\sum_{i=1}^n\left(\boldsymbol{\varphi}_i^{(t-1)} - \sum_{r=1}^R\gamma_{ir}^{(t)}\*B_{r}^{(t)}\boldsymbol{x}_i\right)\boldsymbol{x}_i^\top\right\}\left(\*S_{\omega}^{(t)}\right)^{-1}, \*\Sigma^{(t)}, \left(\*S_{\omega}^{(t)}\right)^{-1}\right);$$
    
    \For{$j \in [d]$}
        \State Sample
        \[
        \nu^{(t)}_j \sim \text{Gamma}\left(\frac{Rq}{2}+a_\nu, \frac{\sum_{r=1}^R\left(\boldsymbol{b}_{jr}^{(t)}\right)^\top\left(\*\Sigma^{(t)}\right)^{-1}\boldsymbol{b}_{jr}^{(t)}}{2}+b_\nu\right),
        \] where $\boldsymbol{b}_{jr}^{(t)}$ is the $j$th column of $\*B_r^{(t)}$;
    \EndFor
    \For{$j \in [d]$}
        \State Sample
        \[
        \omega^{(t)}_j \sim \text{Gamma}\left(\frac{q}{2}+a_\omega, \frac{\left(\boldsymbol{b}_{j0}^{(t)}\right)^\top\left(\*\Sigma^{(t)}\right)^{-1}\boldsymbol{b}_{j0}^{(t)}}{2}+b_\omega\right),
        \] where $\boldsymbol{b}_{j0}^{(t)}$ is the $j$th column of $\*B_0^{(t)}$;
    \EndFor
    \For{$i \in [n]$}
        \State Update 
        \[
        \boldsymbol{\mu}_i^{(t)} \leftarrow \*B_0^{(t)}\boldsymbol{x}_i + \sum_{r=1}^R\gamma^{(t)}_{ir}\*B_r^{(t)}\boldsymbol{x}_i, \quad\*O^{(t)} \leftarrow \left(\*\Sigma^{(t)}\right)^{-1} + \text{diag}\left(\boldsymbol{g}^{(t)}_i\right);
        \]
        \State Sample
        \[
        \boldsymbol{\varphi}_i^{(t)} \sim \sN_q\left(\left(\*O^{(t)}\right)^{-1}\left\{\boldsymbol{y}_i-\frac{\boldsymbol{N}_i}{2}+ \left(\*\Sigma^{(t)}\right)^{-1}\boldsymbol{\mu}_i^{(t)}\right\}, \left(\*O^{(t)}\right)^{-1}\right);
        \]
    \EndFor
    \For{$i \in [n], k \in [q]$}
        \State Sample
        \[
        g_{ik}^{(t)} \sim \text{PolyaGamma}(N_{ik}, \varphi_{ik}).
        \]
    \EndFor
\EndFor
\end{algorithmic}

\section{Details of numerical experiments}
\label{sec:apdx_details_simu}
\subsection{Parameter generation}
\label{subsec:par_gen}

We use a subset of 100 samples in \cite{yatsunenko2012human} to estimate the intercept column of $\*B_0$ and the library sizes $N_{ij}$. These 100 samples were sequenced using both 16S and shotgun metagenomics, although we only used the 16S count data. Distributions of taxa prevalence and sequencing depths in this subset are similar to those of the full dataset used in the data analysis (Section~\ref{sec:real_data}). Before parameter generation, we perform quality control by excluding genera with zero proportion exceeding 0.5 or total counts under 100, resulting in a filtered dataset of 100 samples and 99 genera with a zero-count proportion of 0.18. From this reference set, we randomly sample $q+1$ genera with replacement and transform the leaf-node counts to internal-node observations using the tree-based expression-measurement model \eqref{eq:expression_measurement_2}. For each internal node, the log-odds are calculated to serve as the intercept column for $\*B_0$. Finally, to simulate realistic library sizes, we randomly sample $n$ individuals with replacement and use their total counts at each internal node as the values for $\set{\boldsymbol{N}_i}_{i=1}^n$ in our numerical experiments.

For the population covariance matrix $\*\Sigma$,  we consider three distinct sparsity patterns, all with unit diagonal elements. 
\begin{itemize}
    \item Tridiagonal: the matrix is specified with 0.5 for all first-order off-diagonal elements, $\sigma_{j, j-1} = \sigma_{j-1, j} = 0.5$.
    \item Scale-free: the sparsity pattern is generated using the Barabási-Albert model via preferential attachment with $m=1$ and power = 5. For each edge in the resulting adjacency matrix, we assign a correlation coefficient sampled uniformly from $[-0.9, -0.3] \cup [0.3, 0.9]$. To ensure positive definiteness, we add an identity matrix to the resulting structure and scale the matrix to have unit diagonal elements.
    \item Tree-based: the sparsity pattern is informed by the phylogenetic tree. We define an edge probability between nodes $i$ and $j$ as $P_{ij} = \exp(-d_{ij})$, where $d_{ij}$ is the graph distance of the two nodes on the phylogenetic tree. Non-zero entries are then sampled from a Bernoulli distribution with probability $P_{ij}$. Similar to the scale-free setting, non-zero correlations are sampled uniformly from $[-0.9, -0.3] \cup [0.3, 0.9]$, followed by a diagonal shift and normalization to ensure the matrix is positive-definite with unit diagonal elements.
\end{itemize}

With the above choices of \(\*\Sigma\) and \(\*B_0\), the proportion of zeros in the simulated datasets is around 0.15, which is close to that observed in the Yatsunenko data (post-processing). 


\subsection{Additional results}
\label{subsec:apdx_simu_results}

\begin{figure}[h]
    \centering
    \includegraphics[width=0.95\linewidth]{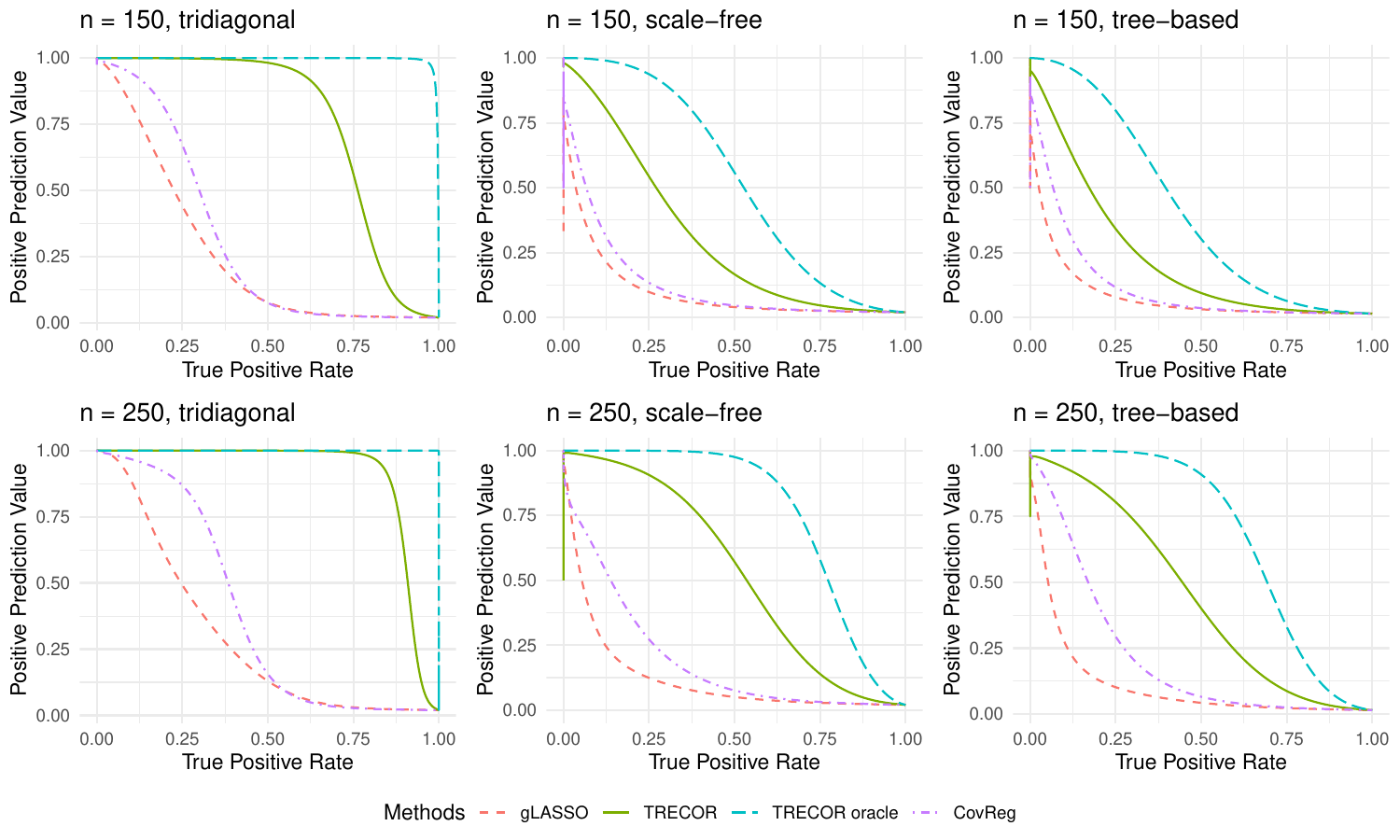}
    \caption{Average precision-recall curves for recovering nonzero off-diagonal entries of the population covariance matrix $\*\Sigma$ across 100 replications. Panels correspond to $n\in\{150,250\}$ and three sparse structures for $\*\Sigma$ (tridiagonal, scale-free, and tree-based). gLASSO (coral dashed), TRECOR (green solid), TRECOR-oracle (light blue longdash) and CovReg (violet dotdash).}
    \label{fig:simu_PR}
\end{figure}

{In addition to ROC curves of Figure~\ref{fig:simu_ROC}, Figure~\ref{fig:simu_PR} reports the average precision-recall curve for recovering nonzero off-diagonal entries of $\*\Sigma$, under the same simulation set-up as described in Section~\ref{sec:simu} and~\ref{subsec:par_gen}. }

\subsection{Impact of zeros}
We conduct numerical experiments to evaluate how the performance of different methods varies with the proportion of zeros in the data. To generate simulated data with different levels of sparsity, we adjust the zero proportion threshold in the quality control step of our parameter generation procedure (see Section~\ref{subsec:par_gen}). This threshold controls the set of genera used to generate the simulated data. By relaxing this threshold, we retain more low-abundance genera, leading to sparser simulated data. In contrast, with a stricter threshold, we retain only genera that are highly abundant, resulting in simulated data with fewer zero entries. We vary the threshold $\in \set{0.3, 0.5, 0.7}$, while keeping the rest of the simulation procedures and evaluation metrics unchanged, as described in Section~\ref{sec:simu} and~\ref{subsec:par_gen}. We focus on the setting where $n = 150$ and $\*\Sigma$ possesses a tree-based structure. 

\begin{figure}[h]
    \centering
    \includegraphics[width=0.95\linewidth]{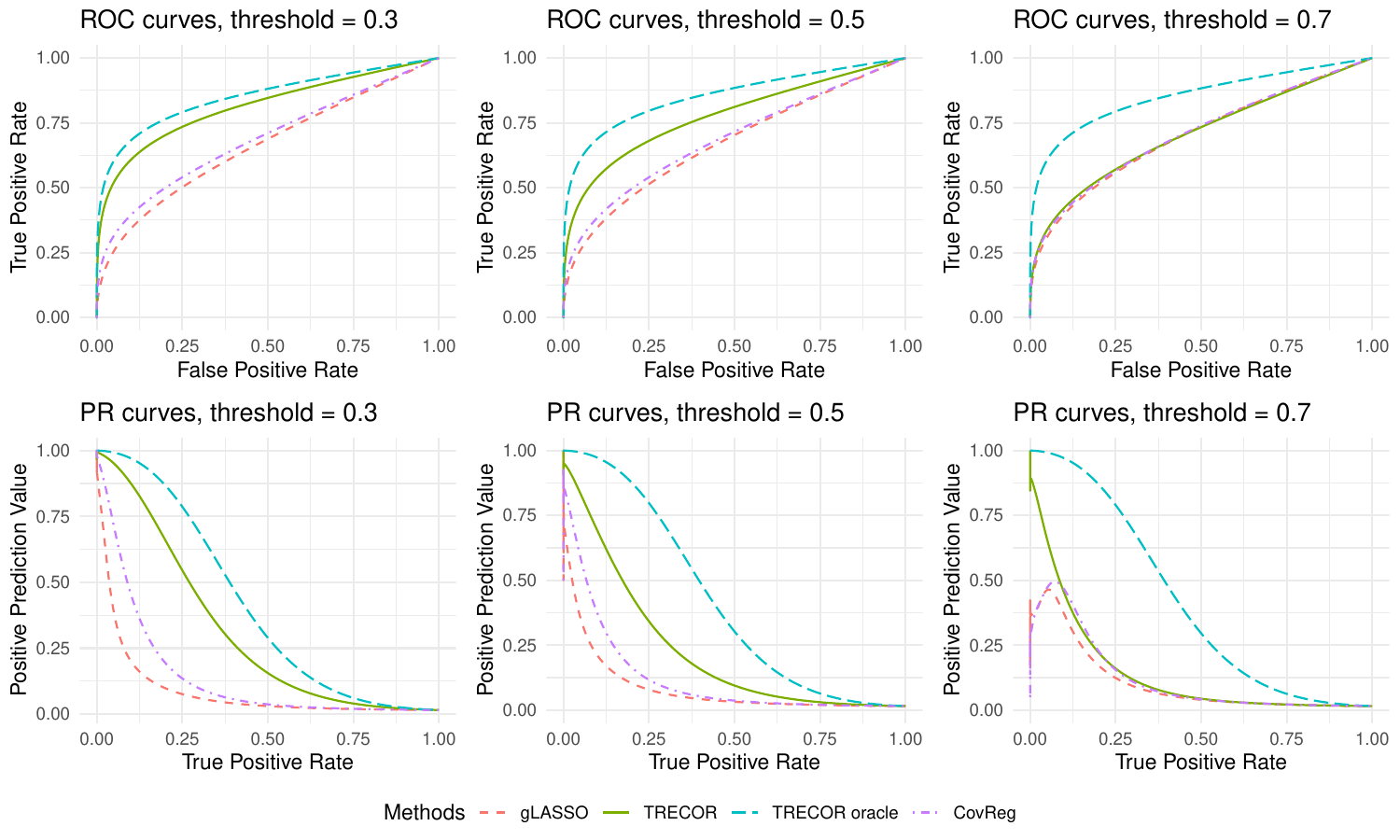}
    \caption{Average ROC and precision-recall curves for recovering nonzero off-diagonal entries of the population covariance matrix $\*\Sigma$ against different sparsity levels of simulated data, averaged over 100 replications. Panels correspond to threshold levels $\in\{0.3, 0.5, 0.7\}$. Curves compare gLASSO (coral dashed), TRECOR (green solid), TRECOR-oracle (light blue longdash) and CovReg (violet dotdash).}
    \label{fig:simu_sparsity}
\end{figure}

Results with the ROC and precision-recall curves for recovering the non-zero elements of $\*\Sigma$, as well as estimation and prediction accuracy, are displayed in Figure~\ref{fig:simu_sparsity} and Table~\ref{tab:simu_sparsity}, respectively. These results show that the performance of all methods except the oracle degrades with increasing levels of sparsity. However, TRECOR uniformly outperforms both gLASSO and CovReg. This demonstrates the importance of properly modeling compositional data. The performance of the oracle method does not vary with the proportion of zeros, because it relies on latent Gaussian variables instead of the observed count data.

\begin{table}[ht]
\centering
\centering
\begin{threeparttable}
\caption{Average estimation and prediction accuracy against different sparsity levels of simulated data, averaged over $100$ replications (standard deviation in parentheses). ``Sparsity" is the average proportion of zeros over 100 simulated datasets. ``Error of $\*\Sigma$'' is the posterior mean squared error for the estimation of $\*\Sigma$. ``Error of $\*\Sigma(\boldsymbol{x})$'' is the prediction error for the estimation of $\*\Sigma(\boldsymbol{x})$. ``Rank selection'' reports the proportion of replicates in which WAIC selects the true rank $R=3$; NA indicates the criterion is not applicable.}
\label{tab:simu_sparsity}
\small
\setlength{\tabcolsep}{5pt}
\renewcommand{\arraystretch}{1.1}
\begin{tabular}{|c|c| l r r r|}
  \hline
 Threshold & Sparsity & Method & Error of $\*\Sigma$ &  Error of $\*\Sigma(\boldsymbol{x})$ & Rank selection\\ 
 \hline
   NA & NA & TRECOR-oracle & 29.56 (0.82) & 2534.08  (293.31) & 0.97\\
 \cline{1-6}
 
 \multirow{3}{*}{0.3} & \multirow{3}{*}{0.06} & gLASSO & 3445.60 (519.83) & 16491.37 (750.30)  & NA \\
  & & TRECOR & 33.57 (1.21) & 2986.44 (315.28) & 0.94  \\
  & & CovReg & 1206.24 (182.74) &  31977.01 (4787.89) & 0.01  \\
  \cline{1-6}

  \multirow{3}{*}{0.5} & \multirow{3}{*}{0.15} & gLASSO & 6564.67  (559.12) & 27413.01 (1236.63) & NA \\
  & & TRECOR & 35.70 (1.60) & 3538.42 (416.22) & 0.81\\
  & & CovReg & 2861.09 (271.01) &  67486.99 (7570.74) & 0.02\\
  \cline{1-6}

  \multirow{3}{*}{0.7} & \multirow{3}{*}{0.30} & gLASSO & 9316.01 (738.70) & 37161.98 (1832.61) & NA \\
  & & TRECOR & 36.89 (2.30) & 5468.17 (429.16) & 0.79 \\
  & & CovReg & 4275.36 (294.59) & 108290.93 (11495.14) & 0.09 \\ 
  \hline
  
\end{tabular}
\end{threeparttable}
\end{table}
\subsection{Choice of hyperparameters}
We conduct numerical experiments to evaluate the sensitivity of the proposed method to different hyperparameter specifications. Following the recommendations of \cite{wang2012bayesian}, we fix $v = 1$ and $s = 0.01$, while varying $a_\nu$, $b_\nu$, $a_\omega$, and $b_\omega$. The simulation framework and evaluation metrics are consistent with those described in Section~\ref{sec:simu} and~\ref{subsec:par_gen}, focusing on the setting where $n = 150$ and $\*\Sigma$ possesses a tree-based structure. 

\begin{figure}
    \centering
    \includegraphics[width=0.95\linewidth]{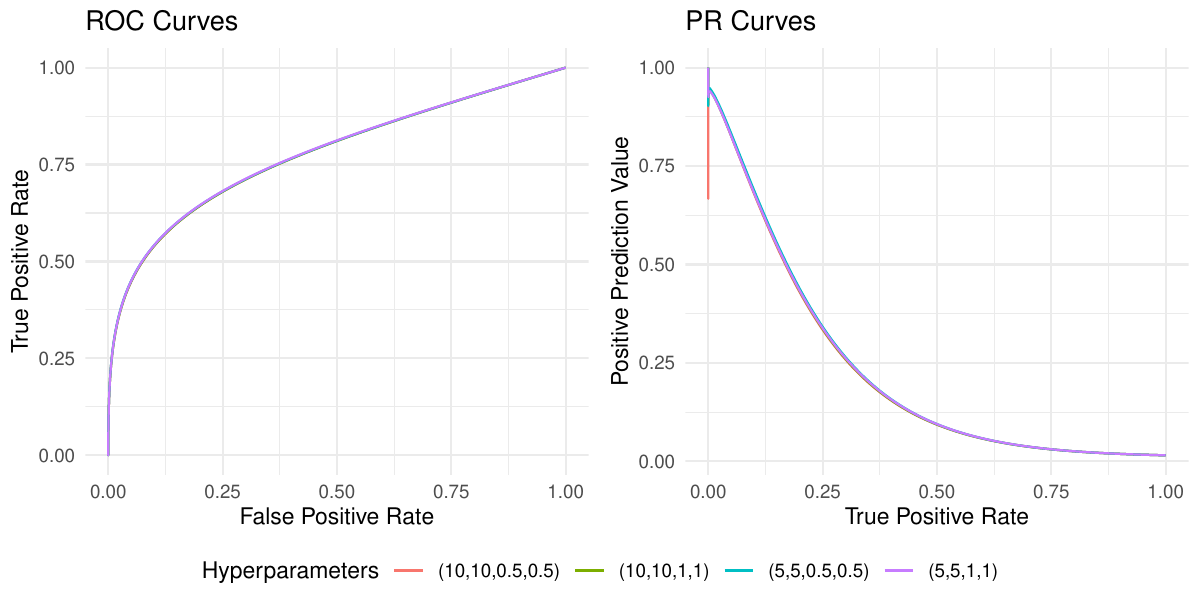}
    \caption{Average ROC curves (left panel) and precision-recall curves (right panel)  for recovering nonzero off-diagonal entries of the population covariance matrix $\*\Sigma$ over different sets of hyperparameters ($a_\nu$, $b_\nu$, $a_\omega$, $b_\omega$). The curves are averaged over 100 simulation replications.}
    \label{fig:hyper_ROC_PR}
\end{figure}

Results with the ROC and precision-recall curve for recovering the non-zero elements of $\*\Sigma$, as well as estimation and prediction accuracy, are displayed in Figure~\ref{fig:hyper_ROC_PR} and Table~\ref{tab:simu_mnse_hyper}, respectively. These results demonstrate that the method is robust, with performance metrics remaining largely invariant across different hyperparameter choices. However, we do note that the hyperparameters $a_\nu$, $b_\nu$, $a_\omega$, and $b_\omega$ determine the amount of $L_2$ penalty imposed on the high-dimensional parameters $\*B_0$ and $\set{\*B_r}_{r=1}^R$. In scenarios where the number of variables $q$ is large relative to the sample size $n$, adopting excessively small values for these hyperparameters may lead to MCMC instability.

\begin{table}[ht]
\centering
\centering
\begin{threeparttable}
\caption{Average estimation and prediction accuracy of TRECOR over different sets of hyperparameters ($(a_\nu$, $a_\omega$, $b_\nu$, $b_\omega)$) averaged over 100 simulation replications (standard deviation in parentheses). ``Error of $\*\Sigma$'' is the posterior mean squared error for the estimation of $\*\Sigma$. ``Error of $\*\Sigma(\boldsymbol{x})$'' is the prediction error for the estimation of $\*\Sigma(\boldsymbol{x})$. ``Rank selection'' reports the proportion of replicates in which WAIC selects the true rank $R=3$.}
\label{tab:simu_mnse_hyper}
\small
\setlength{\tabcolsep}{5pt}
\renewcommand{\arraystretch}{1.1}
\begin{tabular}{l r r r|}
  \hline
 $(a_\nu$, $a_\omega$, $b_\nu$, $b_\omega)$ & Error of $\*\Sigma$ &  Error of $\*\Sigma(\boldsymbol{x})$ & Rank selection\\ 
 \hline
$(5,5,0.5,0.5)$ & 35.70 (1.60) & 3538.42 (416.22) & 0.982\\ 
$(5,5,1,1)$ & 35.22 (1.59) & 3546.29 (350.65) & 1.000\\ 
$(10,10,0.5,0.5)$ & 36.45 (1.68) & 3532.64 (443.48) & 1.000\\ 
$(10,10,1,1)$ & 32.41 (1.00) & 2869.66 (261.42) & 0.987\\ 
   \hline
\end{tabular}
\end{threeparttable}
\end{table}

\section{Details of case study}

\subsection{Exploratory analysis}

\begin{figure}[h!]
    \centering
    \includegraphics[width=0.9\linewidth]{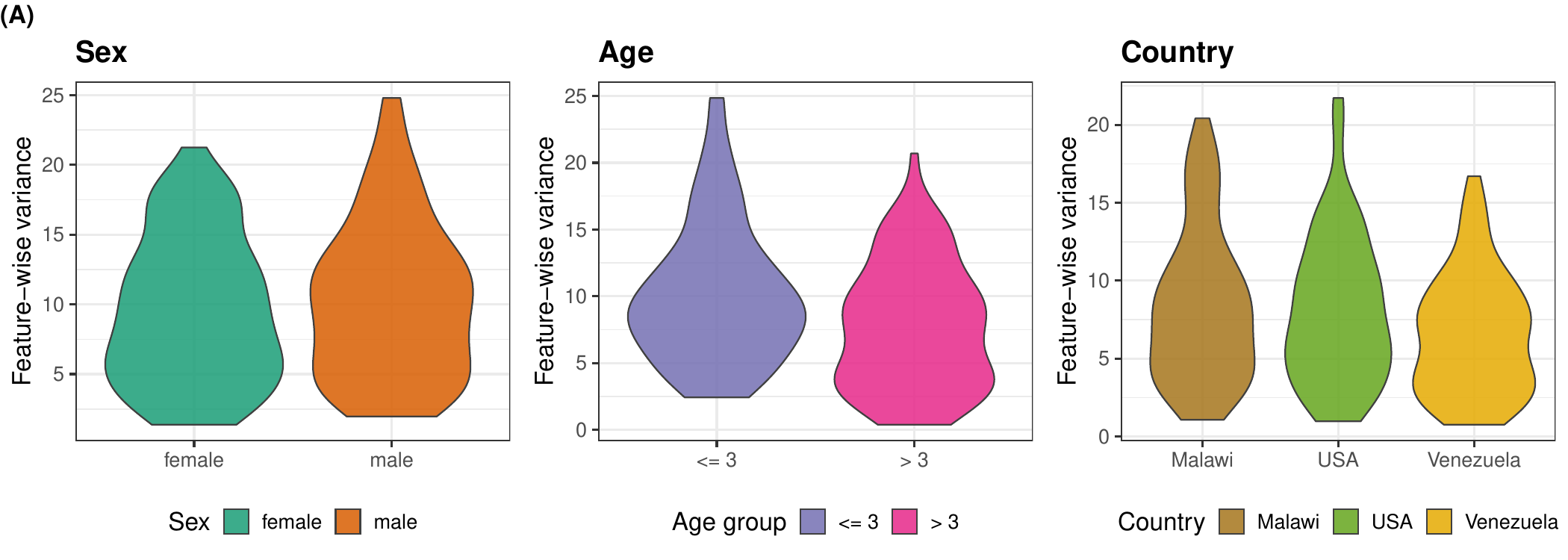}
    \includegraphics[width=0.9\linewidth]{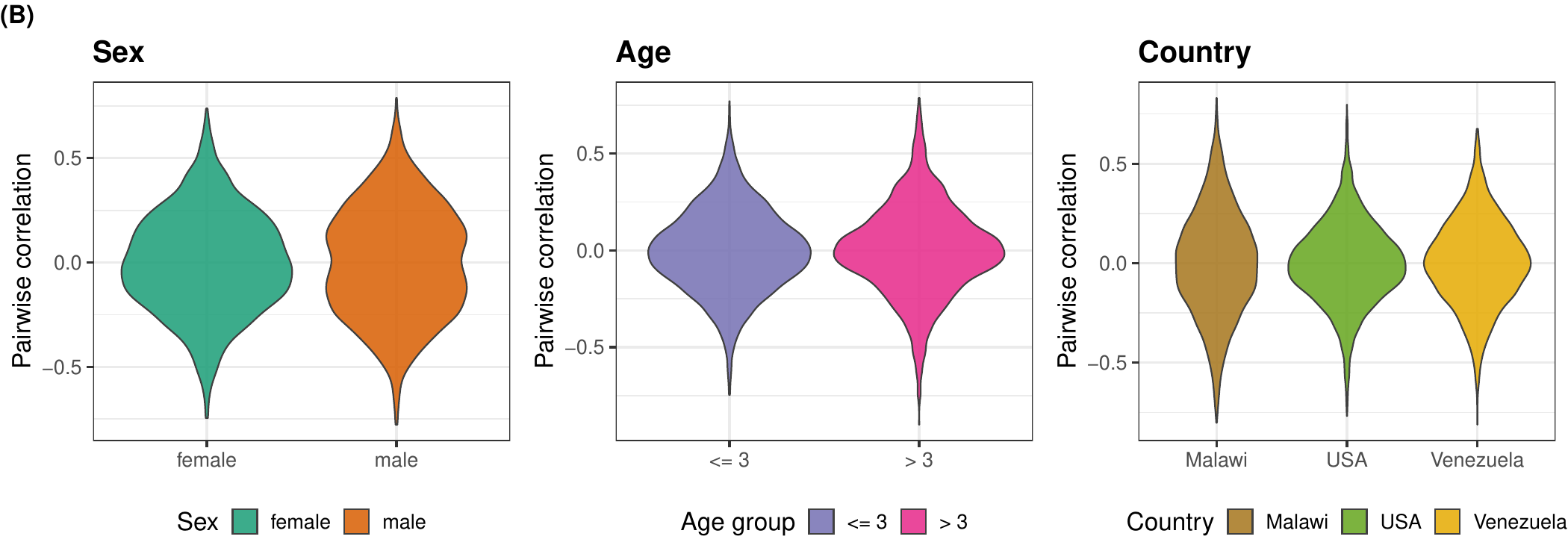}
    \caption{Illustration of covariate-dependent covariation in the \cite{yatsunenko2012human} study. Here features are defined at internal node level. (A) Feature variances; (B) Feature correlations.}
    \label{fig:covariation:internalnodes}
\end{figure}

Figure~\ref{fig:covariation:internalnodes} illustrates the distribution of feature-wise variances and pairwise feature correlations across groups defined by sex, age, and country of origin. Here features are defined as internal nodes on the phylogenetic tree. These patterns are largely consistent with those observed in Figure \ref{fig:covariation}. 

\subsection{Details on constructing the differential correlation network}
\label{subsec:apdx_diff}
Suppose we obtain posterior samples $\*\Sigma^{(l)}$ and $\set{\*B_r^{(l)}}_{r=1}^R$ for $l \in [L]$ after burn-in. For any $\boldsymbol{x} \in \mathbb{R}^d$, define the posterior covariance function at draw $l$ as
$$\*\Sigma^{(l)}(\boldsymbol{x}) = \*\Sigma^{(l)} + \sum_{r=1}^R\*B_r^{(l)}\boldsymbol{x}\boldsymbol{x}^\top\left(\*B_r^{(l)}\right)^\top.$$
Let $\boldsymbol{x}_1$ and $\boldsymbol{x}_2$ be two realizations of the covariate vector. To simplify the expression, let $\mathbf{D}_{i}^{(l)} = \text{diag}(\boldsymbol{\Sigma}^{(l)}(\boldsymbol{x}_i))$ for $i \in \{1, 2\}$ represent the diagonal matrix of variances. 
The posterior differential correlation matrix is defined, for each $l \in [L]$, as
\begin{equation} 
\label{eq:apdx_diff_cor} 
\frac{1}{2} \left[ \big(\mathbf{D}_{1}^{(l)}\big)^{-1/2} \boldsymbol{\Sigma}^{(l)}(\boldsymbol{x}_1) \big(\mathbf{D}_{1}^{(l)}\big)^{-1/2} - \big(\mathbf{D}_{2}^{(l)}\big)^{-1/2} \boldsymbol{\Sigma}^{(l)}(\boldsymbol{x}_2) \big(\mathbf{D}_{2}^{(l)}\big)^{-1/2} \right].
\end{equation}
which is one-half of the difference between the corresponding correlation matrices of $\*\Sigma^{(l)}(\boldsymbol{x}_1)$ and $\*\Sigma^{(l)}(\boldsymbol{x}_2)$.

We next describe the selection of significant entries from the differential correlation matrices. Our procedure follows the Bayesian false discovery rate (FDR) framework of \citet[Section 2.5]{wang2022bayesian}. Let $\Delta_{ij}^{(l)}$ denote the $(i,j)$th entry of \eqref{eq:apdx_diff_cor} for $1 \le i < j \le q$. Fix a threshold $\rho \in (0,1)$ representing the minimum magnitude of differential correlation deemed practically significant. For each pair $(i,j)$, define
$$f_{ij} = \frac{1}{L}\sum_{l=1}^LI\set{\left\lvert\Delta^{(l)}_{ij}\right\rvert \le \rho},$$
which estimates the local Bayesian FDR, i.e., the posterior probability that the differential correlation does not exceed the threshold $\rho$ in magnitude. 
Let $\delta \in (0,1)$ be a target global Bayesian FDR level. Order the values $\set{f_{ij}}_{i<j}$ increasingly as $\set{f^{(k)}_{ij}}_{k=1}^{q(q-1)/2}$. Define $k^*$ as the largest $k \in \left[\frac{q(q-1)}{2}\right]$ such that $$\frac{1}{k}\sum_{k^\prime = 1}^kf_{ij}^{(k)} \le \delta.$$
We then select all pairs $(i,j)$ such that $f_{ij} \le f_{ij}^{(k^*)}$, which controls the global Bayesian FDR at level $\delta$. {Throughout the analysis, we set the minimum magnitude of differential correlation deemed practically significant $\rho = 0.1$ and the global Bayesian FDR $\delta = 0.05$. }

\subsection{Sensitivity analysis and MCMC convergence}
\label{subsec:apdx_convergence}

\begin{figure}[h]
    \centering
    \includegraphics[width=\textwidth]{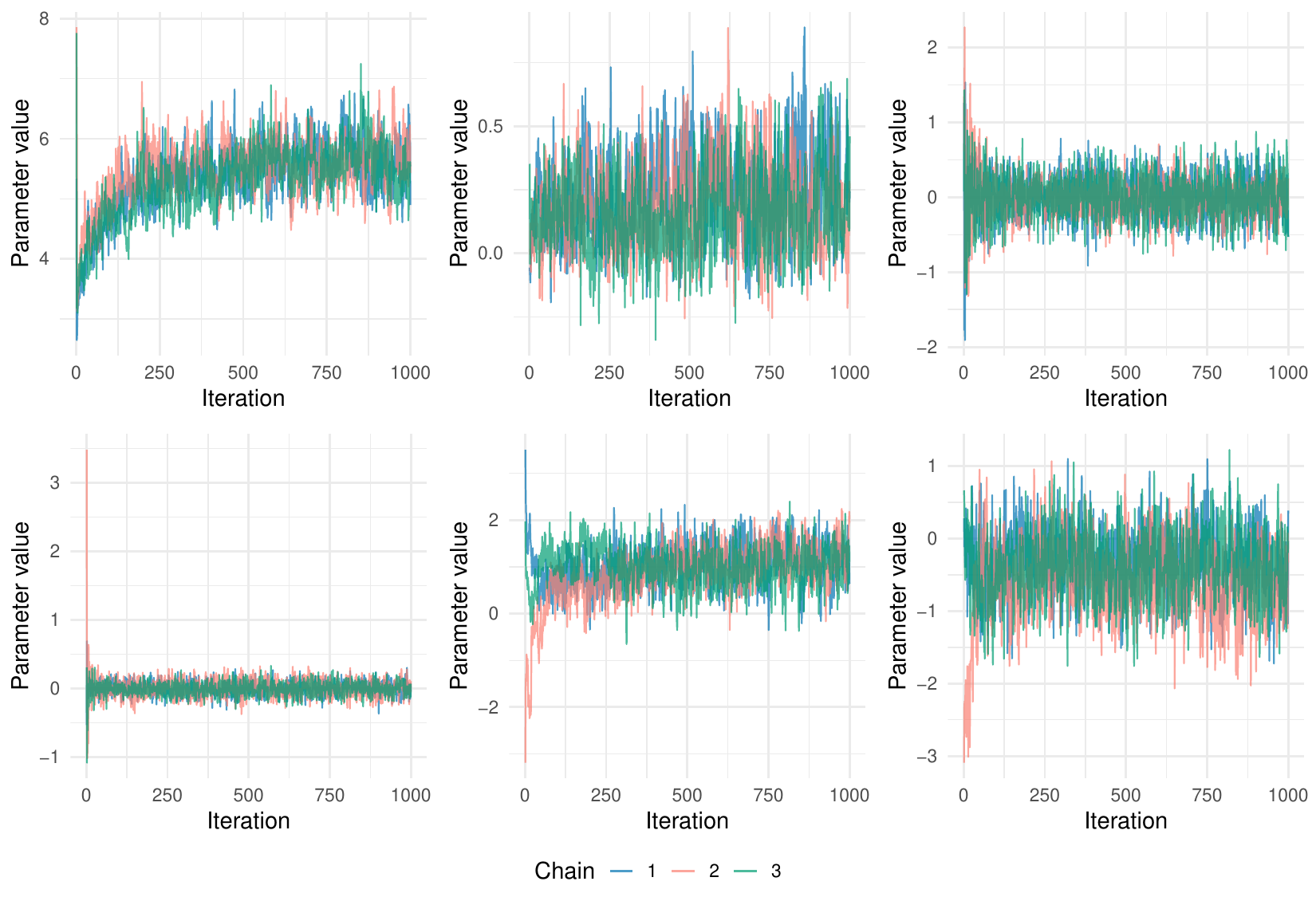} 
    \caption{Trace plots of randomly chosen parameters with three independent Markov chains starting from different initializations.  Only the first $1,000$ iterations of each chain are plotted.}
    \label{fig:MCMC_convergence}
\end{figure}

We assess MCMC convergence using trace plots for a set of randomly selected model parameters. We ran three independent chains initialized at different starting values. The resulting trace plots in Figure~\ref{fig:MCMC_convergence} show good mixing and no apparent nonstationarity, suggesting satisfactory convergence.

\begin{figure}[h]
    \centering
    \includegraphics[width=\textwidth]{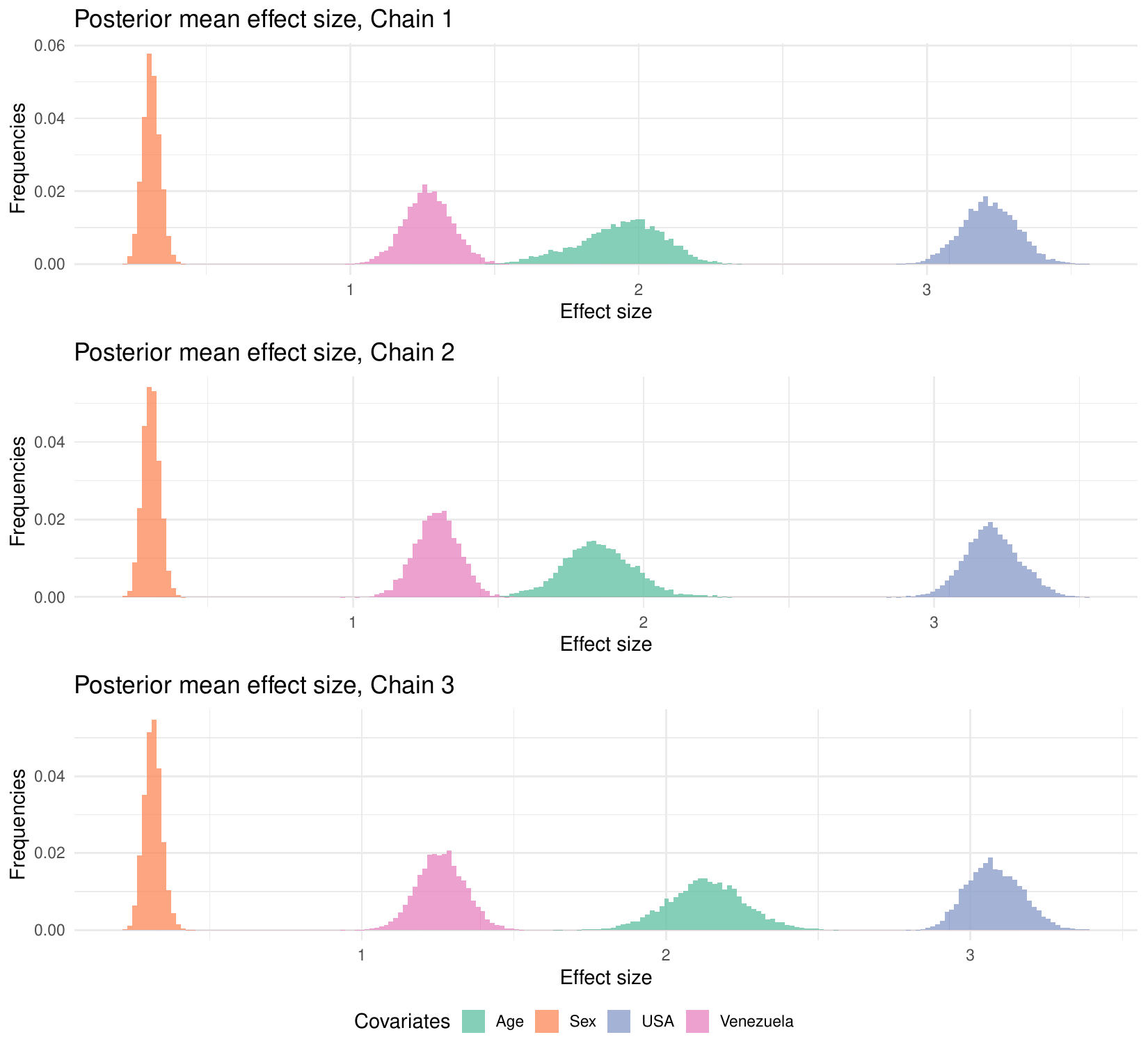} 
    \caption{Posterior mean effect size distribution for each covariate (excluding intercept), represented by the posterior distribution of $\norm{\boldsymbol{b}_{j0}}_2^2$ where $\boldsymbol{b}_{j0} \in \bR^{q}$ is the $j$th column of $\*B_0$, for different MCMC chains.}
    \label{fig:sensitivity_mean_MCMC}
\end{figure}

\begin{figure}[h]
    \centering
    \includegraphics[width=\textwidth]{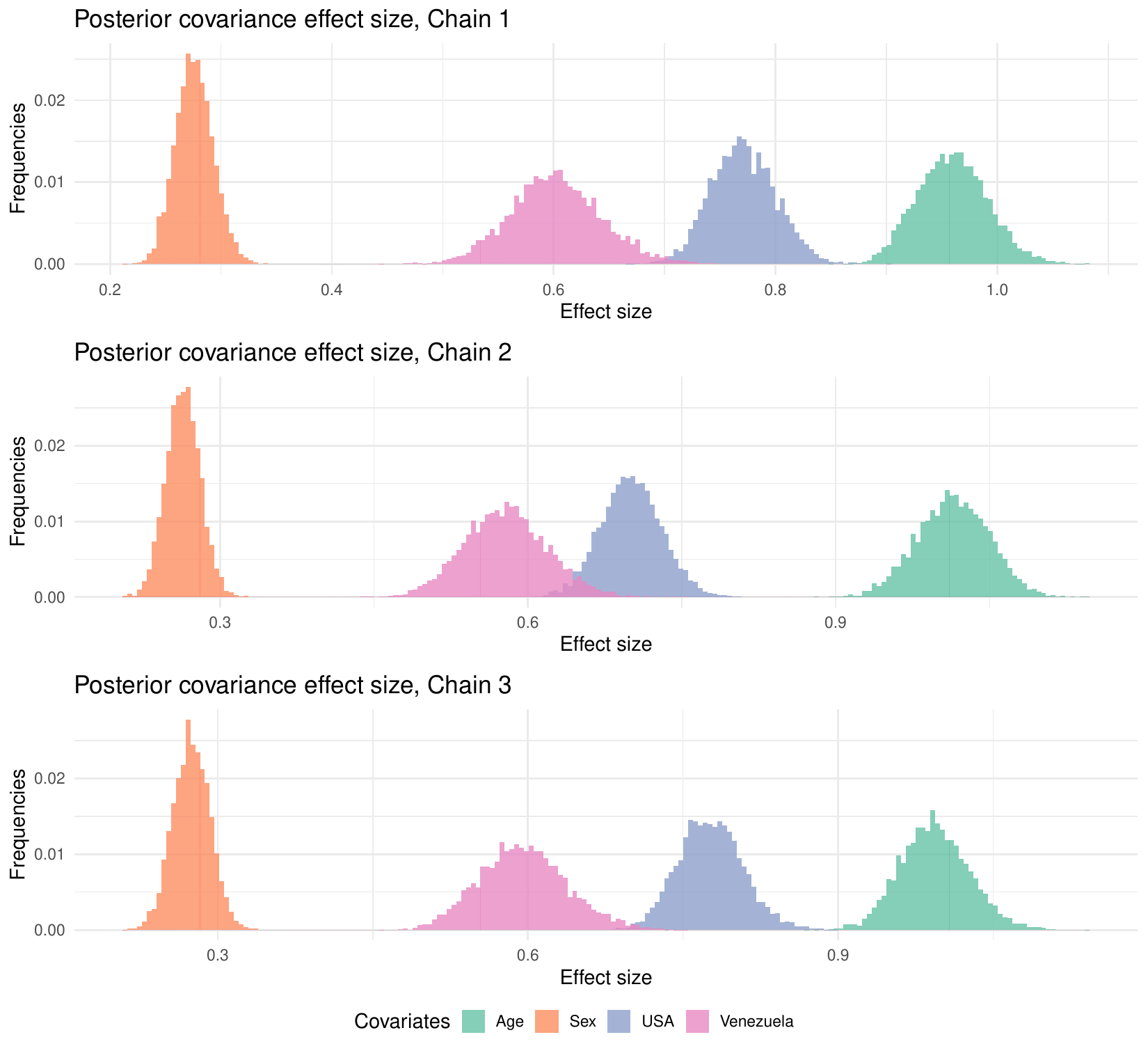} 
    \caption{Posterior covariance effect size distribution for each covariate (excluding intercept), represented by the posterior distribution of $\norm{\*B^{(j)}}_F$, for different MCMC chains.}
    \label{fig:sensitivity_cov_MCMC}
\end{figure}

To evaluate the sensitivity of our findings in Section~\ref{sec:real_data} to different Markov chains, we perform sensitivity analysis by re-running the analysis in Section~\ref{sec:real_data} with different chains. We focus on the analysis for the posterior mean and covariance effect size distributions, following the same procedure in Section~\ref{sec:real_data}. The results, displayed in Figures~\ref{fig:sensitivity_mean_MCMC} and~\ref{fig:sensitivity_cov_MCMC}, are highly consistent across chains, indicating that our conclusions are robust to the different MCMC runs.


\begin{figure}[h]
    \centering
    \includegraphics[width=\textwidth]{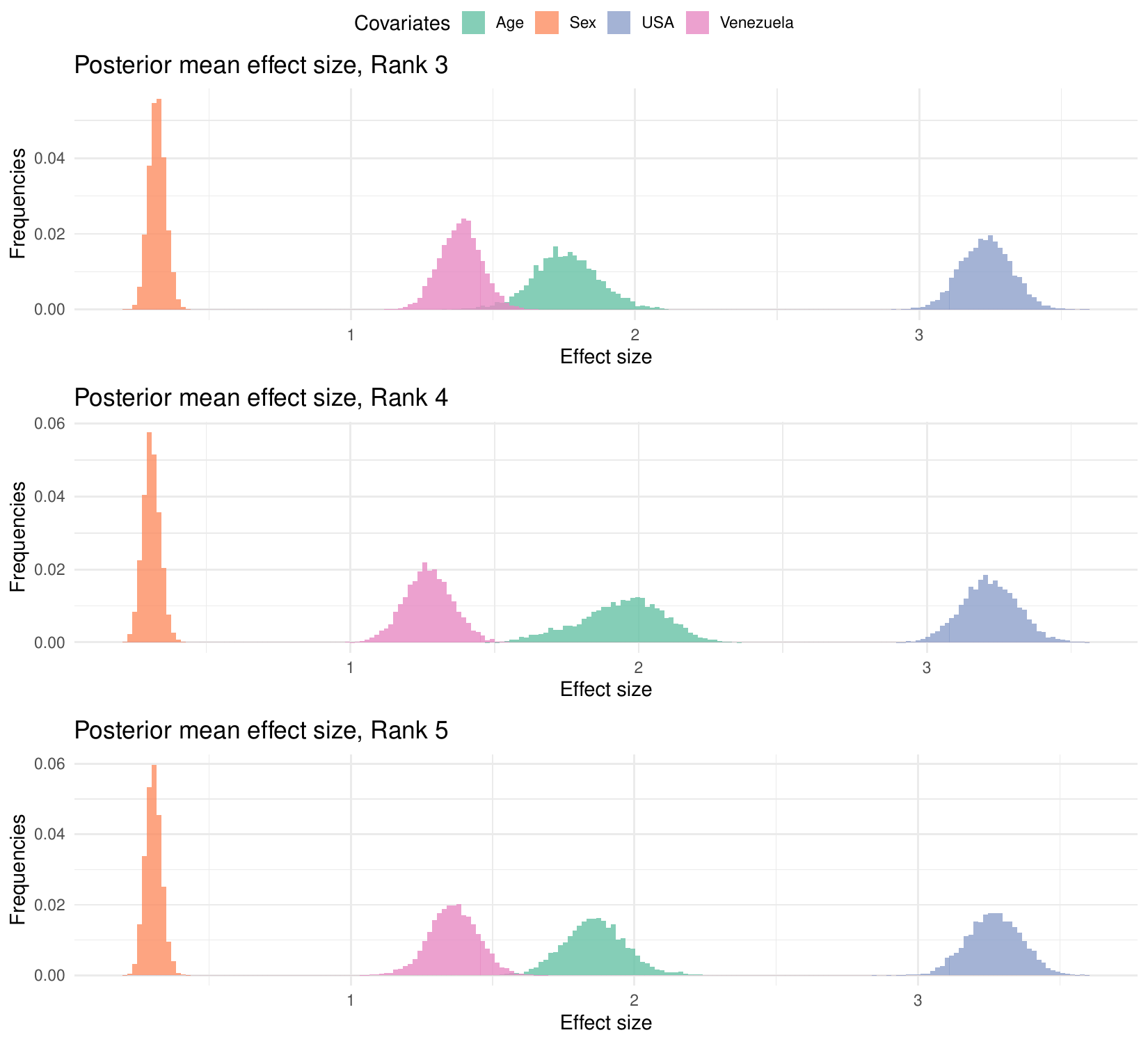} 
    \caption{Posterior mean effect size distribution for each covariate (excluding intercept), represented by the posterior distribution of $\norm{\boldsymbol{b}_{j0}}_2^2$ where $\boldsymbol{b}_{j0} \in \bR^{q}$ is the $j$th column of $\*B_0$, for different rank parameters $R \in \set{3,4,5}$.}
    \label{fig:sensitivity_mean}
\end{figure}

\begin{figure}[h]
    \centering
    \includegraphics[width=\textwidth]{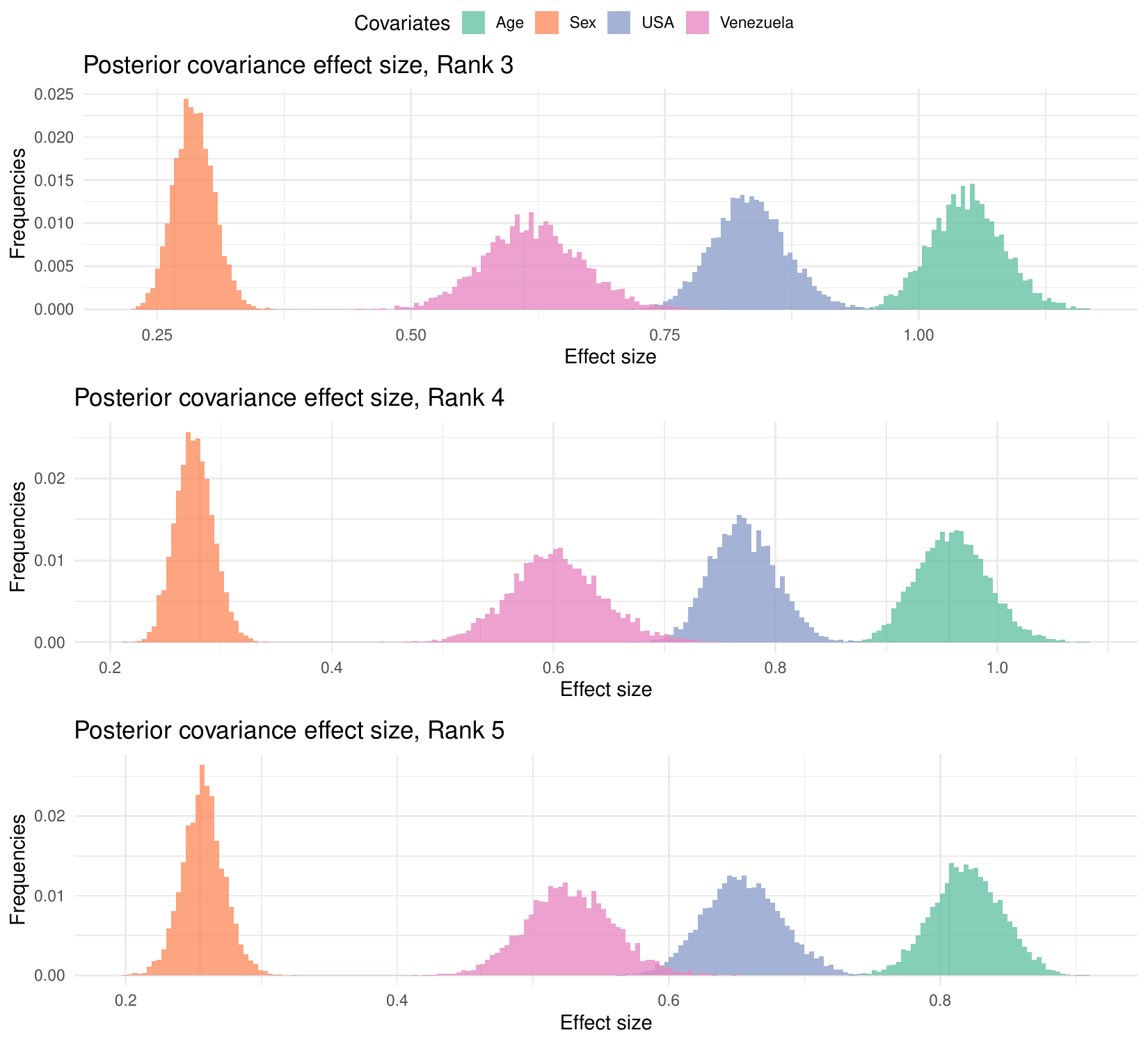} 
    \caption{Posterior covariance effect size distribution for each covariate (excluding intercept), represented by the posterior distribution of $\norm{\*B^{(j)}}_F$, for different rank parameters $R \in \set{3,4,5}$.}
    \label{fig:sensitivity_cov}
\end{figure}

We further perform sensitivity analysis with regard to the rank parameter $R$. In Section~\ref{sec:real_data}, we use $R=4$ as selected by WAIC. Similarly, we re-run the analysis of posterior mean and covariance effect size distributions with $R \in \set{3,4,5}$. 
The results are shown in Figure~\ref{fig:sensitivity_mean} and Figure~\ref{fig:sensitivity_cov}, which are similar across different $R$, suggesting that our results are robust to the rank parameter chosen. 
\end{document}